\documentclass[preprint]{emulateapj}

\usepackage{amsmath}
\usepackage{amssymb}
\usepackage{bm}
\usepackage{ulem}
\usepackage[table]{xcolor}
\usepackage{xspace}
\colorlet{tablegray}{gray!25}

\usepackage{color}

\def\suzaku{{\it Suzaku}\xspace}
\def\vela{{Vela X-1}\xspace}
\defcitealias{Odaka:2013}{Paper I}

\slugcomment{}

\shorttitle{}
\shortauthors{Odaka et al.}

\begin{document}


\title{Short-Term Variability of X-rays from Accreting Neutron Star Vela X-1: II. Monte-Carlo Modeling}


\author{Hirokazu Odaka\altaffilmark{1}, Dmitry Khangulyan\altaffilmark{1}, Yasuyuki T.\ Tanaka\altaffilmark{2}, Shin Watanabe\altaffilmark{1}, \\  Tadayuki Takahashi\altaffilmark{1,3}, and Kazuo Makishima\altaffilmark{3,4}}


\altaffiltext{1}{Institute of Space and Astronautical Science (ISAS), Japan Aerospace Exploration Agency (JAXA), 3-1-1 Yoshinodai, Chuo, Sagamihara, Kanagawa, 252-5210, Japan}
\altaffiltext{2}{Hiroshima Astrophysical Science Center, Hiroshima University, 1-3-1 Kagamiyama, Higashi-Hiroshima, Hiroshima 739-8526, Japan}
\altaffiltext{3}{Department of Physics, University of Tokyo, 7-3-1 Hongo, Bunkyo, Tokyo, 113-0033, Japan}
\altaffiltext{4}{The Institute of Physical and Chemical Research, 2-1 Hirosawa, Wako, Saitama, 351-0198, Japan}

\newcommand{\degree}{$^\circ$~}


\begin{abstract}
We develop a Monte Carlo Comptonization model for the X-ray spectrum
of accretion-powered pulsars. Simple, spherical, thermal Comptonization
models give harder spectra for higher optical depth, while the
observational data from Vela X-1 show that the spectra are harder at
higher luminosity. This suggests a physical interpretation where the
optical depth of the accreting plasma increases with mass accretion
rate. We develop a detailed Monte-Carlo model of the accretion flow,
including the effects of the strong magnetic field ($\sim 10^{12}$~G)
both in geometrically constraining the flow into an accretion column,
and in reducing the cross section. We treat bulk-motion Comptonization
of the infalling material as well as thermal Comptonization. These model spectra
can match the observed broad-band {\it Suzaku} data from Vela X-1
over a wide range of mass accretion rates. The model can also explain
the so-called ``low state'', in which the luminosity decreases by an
order of magnitude. Here, thermal Comptonization should be negligible,
so the spectrum instead is dominated by bulk-motion Comptonization.
\end{abstract}




\bibliographystyle{apj}


\section{Introduction}\label{sec:discussion_comptonization}
Binary systems harboring a massive star and a black hole (or a neutron star) form the population of the brightest X-ray sources. In these sources emission is powered by accretion from the optical companion onto the compact object, and is typically dominated by the thermal emission from the accretion disk. However, this is not the case when the compact object is a strongly magnetized neutron star (i.e., a pulsar). The key difference between this and the conventional scenario is related to the disruption of the accretion flow by the pulsar magnetic field. The structure  of the disrupted flow is governed by the magnetic field, which re-directs the accreted plasma towards the magnetic poles of the pulsar. This results in the formation of a so-called {\it accretion column} \citep[see, e.g.,][and references therein]{Becker:2005}. The physical conditions realized in the accretion column differ significantly from ones typically formed in accretion disks. Therefore, accreting binary pulsars represent a subclass of bright X-ray sources with rather  specific properties. Given the complexity of the processes taking place in the accretion column, a proper interpretation of observational data obtained in the X-ray energy band requires a detailed modeling of an entire range of different physical processes. 

Conservation of energy implies that the {\bf luminosity} is roughly
determined by the accretion rate. However, the spectral shape does not allow any simple interpretation.  Therefore, studying of the spectral properties might shed light on the physical conditions of the accretion flow in binary pulsars.  \vela is the brightest wind-fed accreting pulsar and displays strong time variability in the X-ray band \citep[e.g.][]{Kreykenbohm:2008}, therefore this object provides us with an ideal laboratory for studying the physical conditions of the accreted plasma.  In order to obtain information regarding spectral features of the emission produced by the accretion flow in \vela, we analyzed X-ray data collected with \suzaku \citep[][hereafter \citetalias{Odaka:2013}]{Odaka:2013}.
The observation was 145 ks long and has an exposure time of 100 ks.
Owing to the wide-band coverage and the high signal-to-noise performance of instruments onboard \suzaku \citep{Mitsuda2007,Takahashi:2007,Kokubun:2007}, we extracted 56 sequential X-ray spectra of Vela X-1 for short time intervals of 2 ks in an energy range from 2.5 keV to 50 keV.
These spectra sample a wide luminosity range, allowing us to constrain the physical parameters of the accretion flow.

As it was shown in \citetalias{Odaka:2013}, the X-ray spectra obtained with \suzaku from
\vela are the best fitted by the ``NPEX'' model \citep[a combination of negative and positive power laws with an
exponential cutoff by a common folding energy;][]{Mihara:1995, Makishima:1999}, i.e., by
\begin{equation}\label{eq:model_npex}
\frac{dN}{dE} = (A_1E^{-\Gamma}+A_2E^{2})\exp\left(-\dfrac{E}{E_f}\right)\,,
\end{equation}
where $A_1$, $A_2$, $\Gamma>0$ and $E_f$ are parameters used for fitting. However, the relation of these parameters to the physical properties of the emitter is vague and, obviously, can vary with a specific physical scenario. This ambiguity is simply a reflection of the complexity of radiation mechanisms operating in these sources.

In this work, we attempt to link the physical properties of the flow to its observational manifestation as seen in the X-ray  energy band from \vela \citepalias[obtained in][]{Odaka:2013}. We consider Comptonization to be the dominant process for X-ray production. To account for a few key processes, in particular, for the   geometry of accretion column,  we exploited  the Monte-Carlo approach. The paper is organized as follows: in Section 2 we present a description of the structure of the accretion flow together with simple analytical estimates, in Section 3 we present results of computations of pure thermal Comptonization, in Section 4 we report results obtained for the assumed structure of an accretion column. Finally, in Section \ref{sec:discussion} we discuss our results and in Section 6 we present our concussions.

\section{Accretion flow in \vela}
For the conditions expected in \vela, the optical star likely does not fill its Roche lobe. Therefore,  the compact object can only accrete  stellar matter via direct wind accretion. In a simple picture of this wind-fed accretion, the accretion flow is described by an analytical solution \citep{Bondi:1944}, which allows a straight forward interpretation: a small fraction of  the wind material (wind passing within the so-called accretion radius of the compact object) is directly captured by the compact star's gravity, whereas the wind outside the radius escapes. 
%
Since the gravitational energy of the accreted matter is converted into X-ray radiation, this approach allows us to obtain an estimate of the 
 X-ray luminosity $L_X$ of the source:
\begin{equation}\label{eq:luminosity_wind_fed_accretion}
L_{\rm X} =  \frac{(GM_*)^3\dot{M}_\mathrm{wind}}{R_* v_\mathrm{wind}{}^4 D^2} \sim 10^{36}\ \mathrm{erg\ s^{-1}},
\end{equation}
where $R_*\sim10$~km, $v_\mathrm{wind}\sim1000\rm \, km\,s^{-1}$, $M_*\sim1.9M_\odot$, $\dot{M}_\mathrm{wind}\sim10^{-6}M_\odot\mathrm{yr}^{-1}$ and $D\sim50R_\odot$ are the radius of the accreting neutron star; the relative velocity of the wind seen from the accretion center; the mass of the compact star; the mass-loss rate of the donor star by the stellar wind; and  the distance of the compact object from the center of the companion (donor) star, respectively \citep[the values used for normalization of Equation~\eqref{eq:luminosity_wind_fed_accretion} are similar to those expected in \vela, see][and references therein]{Watanabe:2006}.  Despite large uncertainties, Equation \eqref{eq:luminosity_wind_fed_accretion} provides a good estimate for typical X-ray luminosity of \vela.

\begin{figure}[tbp]
\begin{center}
\includegraphics[width=7.0cm]{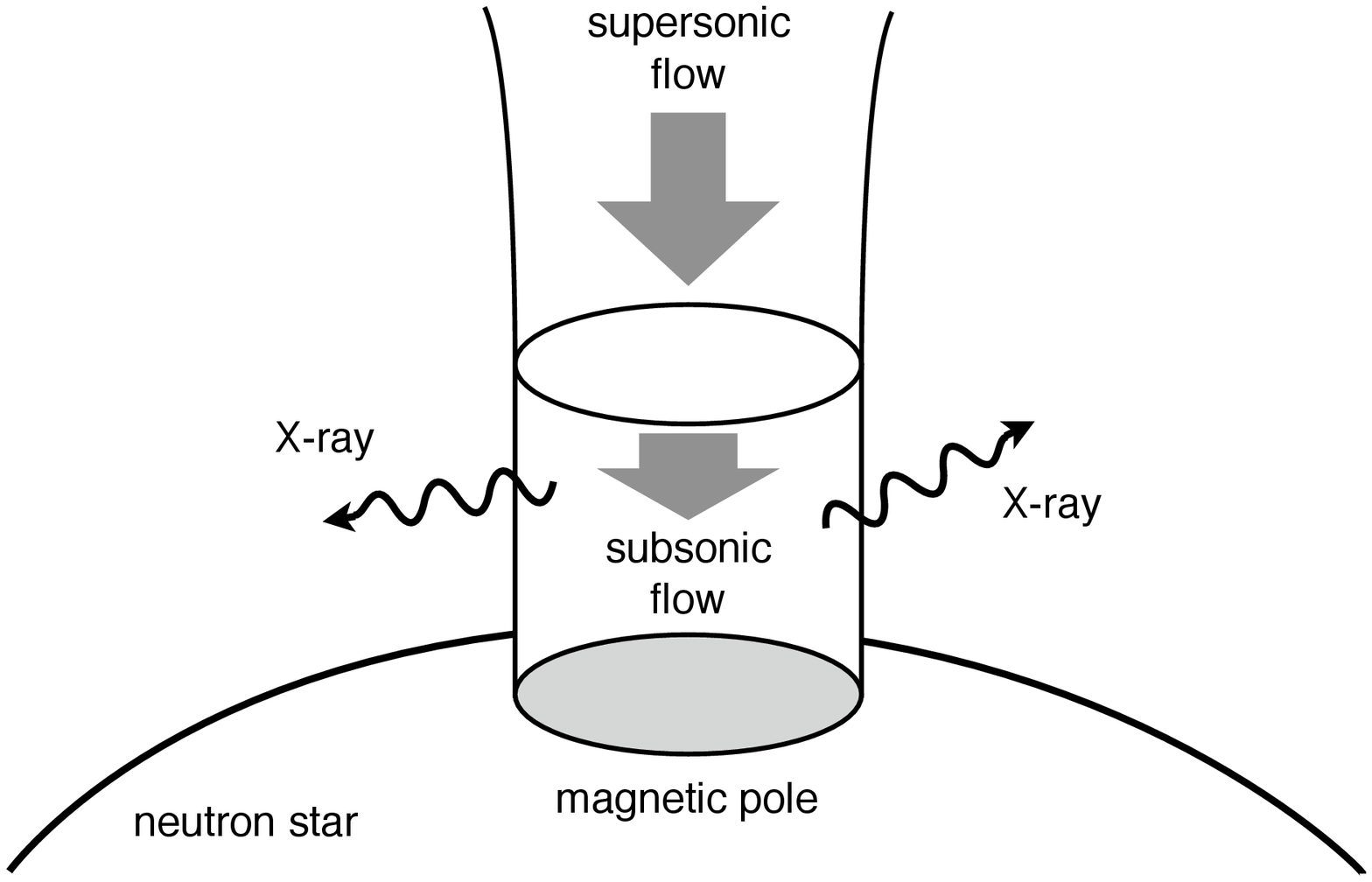}
\includegraphics[width=4.0cm]{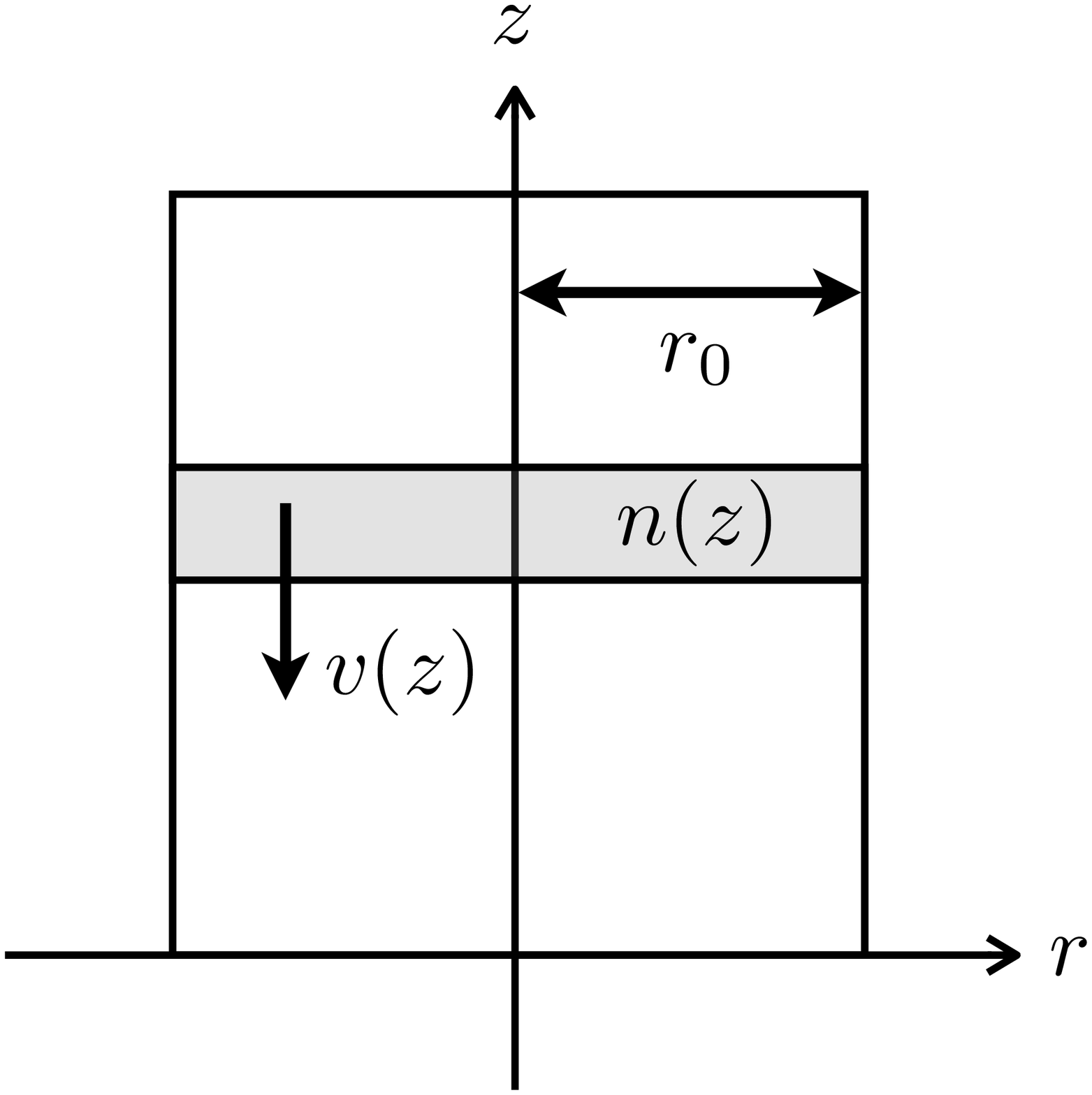}
\caption{Schematic picture of an accretion column on a neutron star (upper panel); and sketch of its structure adopted for numerical modeling (bottom panel).}
\label{fig:accretion_column}
\label{fig:column_coordinates}
\end{center}
\end{figure}

In contrast to the total luminosity, the emission spectral shape is not constrained from the first principles and is strongly influenced by the dominant radiation process, conditions of the accreted plasma and geometry of the flow. In the case of accretion onto a magnetized neutron star, the presence of the magnetic field should strongly influence all these properties.  Indeed, close to the neutron star the magnetic field rapidly increases as $B=\mu/R^3$ (here $\mu=B_* R_{*}^{3}\sim10^{30}\rm \,G\,cm^3$; $B_*$ denotes the magnetic field at the stellar surface), and for conditions expected in a wind-fed binary pulsar, the magnetic pressure should exceed the accretion flow ram pressure at certain distance, $R_{\rm M}$ (known as Alfv\'{e}n radius). Accounting for the typical values of the relevant parameters and  given the relatively weak dependence on them, the Alfv\'{e}n radius is expected to be two or three orders of magnitude larger than the size of the neutron star \citep[see e.g.][\S 6.5]{Frank:2002}. Thus, in the conventional scenario, the strong magnetic field of the neutron star disrupts the accretion flow at $R_{\rm M}$, and then channels it along the field lines to magnetic poles on the neutron star \citep{Basko:1976}. The flow forms a column-like structure, which is called accretion column, as shown in Figure~\ref{fig:accretion_column}. 

It is obvious that the flow moving at free-fall velocity, $v_\mathrm{ff} = \sqrt{2GM_*/R_*}\sim 0.5c$, cannot be accreted by the neutron star. So, the flow is to be decelerated, and the excess of energy should be radiated through some mechanism. Thus, the dominant emission component should be formed in this compact region with quite specific physical properties.  However, currently there is no self-consistent theory, which explains the dynamical and radiative properties of the accretion column. Such a theory should address the extreme complexity of nonlinear physical processes when the gas dynamics is completely coupled to both strong magnetic and strong radiation fields.

For high-luminosity X-ray pulsars, which are characterized by luminosities of $\sim 10^{38}\ \mathrm{erg\ s^{-1}}$, the above scenario was shown to be quite feasible  \citep[see, e.g.,][]{Becker:1998}. Indeed, the radiation pressure in such systems is strong enough to form a radiation-dominated shock, which decelerates the accretion flow. While the in-falling gas passes through the shock, the kinetic energy is converted into X-rays and escapes from the accretion column. The critical luminosity for formation of a radiation-dominated shock is
\begin{equation}
L_\text{crit} = 3\times 10^{37}\left(\frac{r_0}{R_*}\right) \left(\frac{M_*}{M_\odot}\right) \mathrm{erg\ s^{-1}},
\end{equation}
where $r_0$ denotes the radius of the accretion column \citep{Basko:1976, Becker:1998}. Though wind-fed pulsars, such as \vela, have relatively low luminosities $L_X \sim 10^{36}\ \mathrm{erg\ s^{-1}}$ due to smaller accretion rates, it is likely that they also possess radiation-dominated shocks close to the neutron star surfaces \citep{White:1983, Becker:2007}.


In what follows we adopted a one-dimensional model of the flow in the accretion column \citep{Becker:2007}, i.e., all the plasma characteristics depend on only one coordinate: distance to the surface of neutron star ($z$ coordinate), and the magnetic field is assumed to have only $z$-component (see Figure~\ref{fig:column_coordinates}). In this model, the flow is decelerated by radiation-dominated shock near the surface, and the velocity profile derived by \citet{Becker:1998} can be written as
\begin{equation}\label{eq:column_velocity_profile}
v(z) = -v_\mathrm{ff}\left[ 1-\left(\frac{7}{3}\right)^{-z/z_\mathrm{sp}} \right]\,.
\end{equation}
Here, $v_\mathrm{ff}$ denotes the free-fall velocity, and the sonic point $z_\mathrm{sp}$, at which the velocity of the accretion flow becomes equal to a speed of sound, are given by
\begin{equation}\label{eq:sonic_point}
z_\mathrm{sp} = \frac{r_0}{2\sqrt{3}}\left( \frac{\sigma_\perp}{\sigma_\parallel} \right)^{1/2}\ln\frac{7}{3} = 0.245\ r_0 \left( \frac{\sigma_\perp}{\sigma_\parallel} \right)^{1/2}\,,
\end{equation}
where $\sigma_{\parallel/\perp}$ are energy-averaged scattering cross sections for photons propagating parallel and perpendicular to the magnetic field, respectively.
We can also obtain the number density of the plasma as a function of
$z$ by mass conservation (see Equation~\ref{eq:mdot_column}).  In the case of
the radiation-dominated shock, the velocity and density profiles are
continuous unlike ``normal'' gas-dominated shocks, which have a
discontinuous transition at the sonic point.


\section{Thermal Comptonization in Spherical Model}

Although the fast bulk motion and the strong magnetic field affect the X-ray radiation from the accretion column, it is worthwhile to
consider a simple, spherical model without magnetic fields and bulk motions prior to investigating more realistic but more complicated columnar flow models in Section \ref{sec:accretion_column_model}.
We therefore first examine the observed spectral shapes of \vela in a
framework of a pure thermal Comptonization model. Using the Monte Carlo framework MONACO \citep{Odaka:2011}, we performed simulations of spherical uniform Comptonizing clouds with different radial
optical thicknesses.  Based on  this simple approach we estimate characteristic, effective
parameters of the accretion plasma in the X-ray emitter.

\subsection{Comptonization Spectrum}\label{subsec:discussion_comptonization_spectrum}

A Comptonization spectrum mainly depends on three characteristics of the source: the spectrum of the seed photons, the energy distribution of the Comptonizing electrons, and the optical depth. Since the X-ray cutoff energy provides a rough estimate of the electron temperature $kT$, the spectral shape
revealed in \citetalias{Odaka:2013} (see \S~4.2) indicates that the plasma temperature should be close to $kT\sim7\rm keV$. We
therefore consider two cases for temperature, $kT=6\ \mathrm{keV},\ 10\ \mathrm{keV}$, in the simulations.  The electron temperature also allows a natural definition of the spectrum of the seed photons. Namely, the
energy spectrum of the seed photons has been assumed to be thermal bremsstrahlung from a plasma with the same temperature as the
Comptonizing electrons.  More specifically, the photon number distribution of the seed photons is
\begin{equation}\label{eq:source_function_freefree}
\begin{split}
N^\mathrm{ff}(E)&=\frac{dn}{dVdtdE}=\frac{\varepsilon_\nu^\mathrm{ff}}{hE} \\
&= 3.0\times 10^{-15}\ \left(\frac{E}{1\ \mathrm{keV}}\right)^{-1}\left(\frac{kT}{1\ \mathrm{keV}}\right)^{-1/2}\\
&\quad \times Z^2n_\mathrm{e}n_\mathrm{i}e^{-E/kT}\bar{g}_\mathrm{ff} \ \mathrm{photons\ cm^{-3}\ s^{-1}\ keV^{-1}},
\end{split}
\end{equation}
where $\varepsilon_\nu^\mathrm{ff}$ is the energy emissivity of thermal bremsstrahlung; $n_\mathrm{e}$ and $n_\mathrm{i}$ are number densities of electrons and ions, and $\bar{g}_\mathrm{ff}$ is a velocity-averaged Gaunt factor \citep{Rybicki:radiative_process}.

In the numerical computations described in this section we assumed that the source of the seed photons is located in the center of electron cloud.
This setup of the photon source is not the case of a realistic thermal, Comptonizing plasma, in which seed photons can be generated anywhere via bremsstrahlung by the plasma itself.
However, we put the source at the center in order to characterize the thermal Compton spectrum
by the optical depth seen by a photon traveling from the center toward the surface of the cloud. 
This characteristic value can be regarded as an ``effective'' optical depth that indicates a degree of thermal Comptonization.

\begin{figure}[tbp]
\begin{center}
\includegraphics[width=8.5cm]{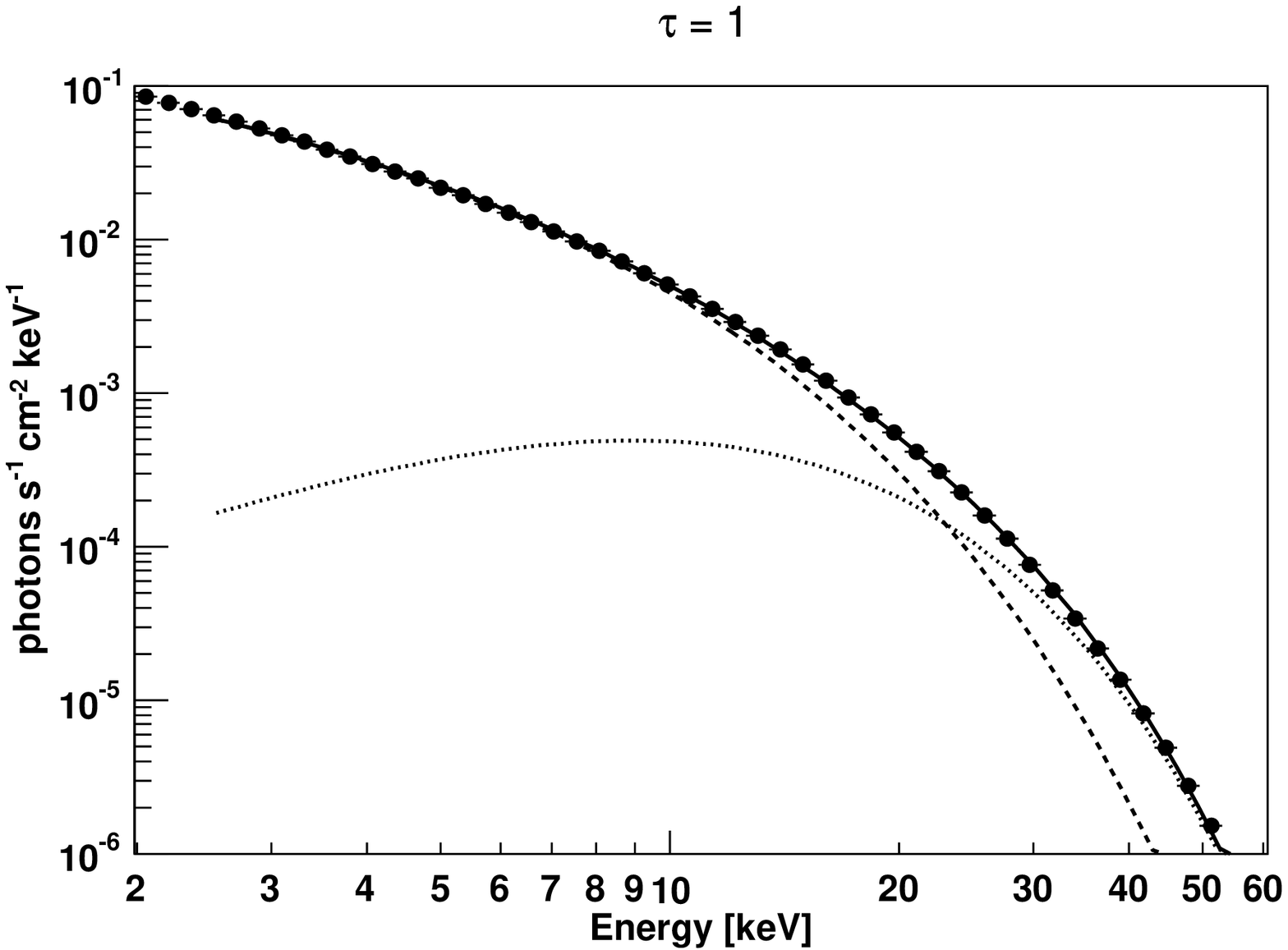}\\
\vspace{12pt}
\includegraphics[width=8.5cm]{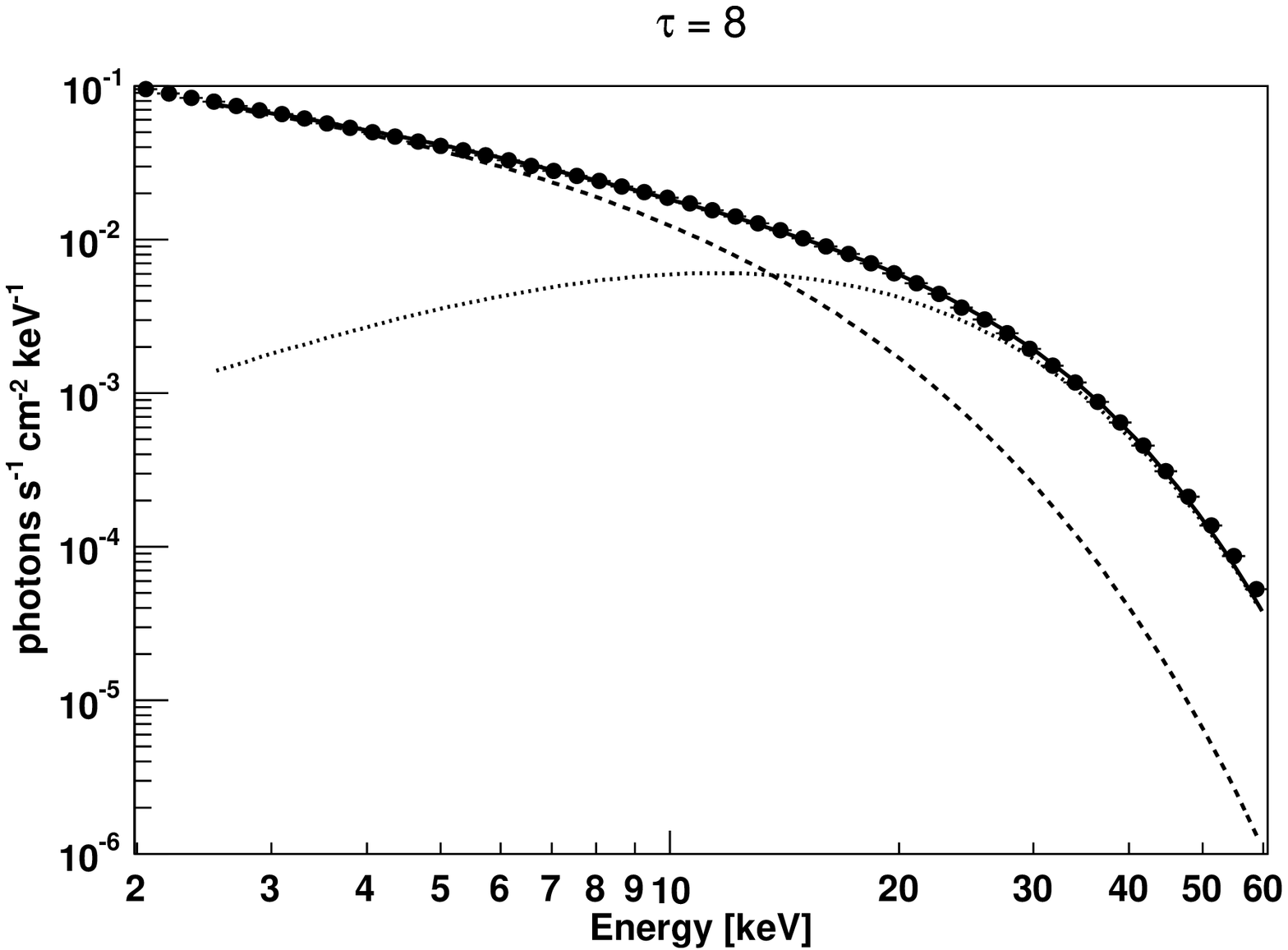}\\
\vspace{12pt}
\includegraphics[width=8.5cm]{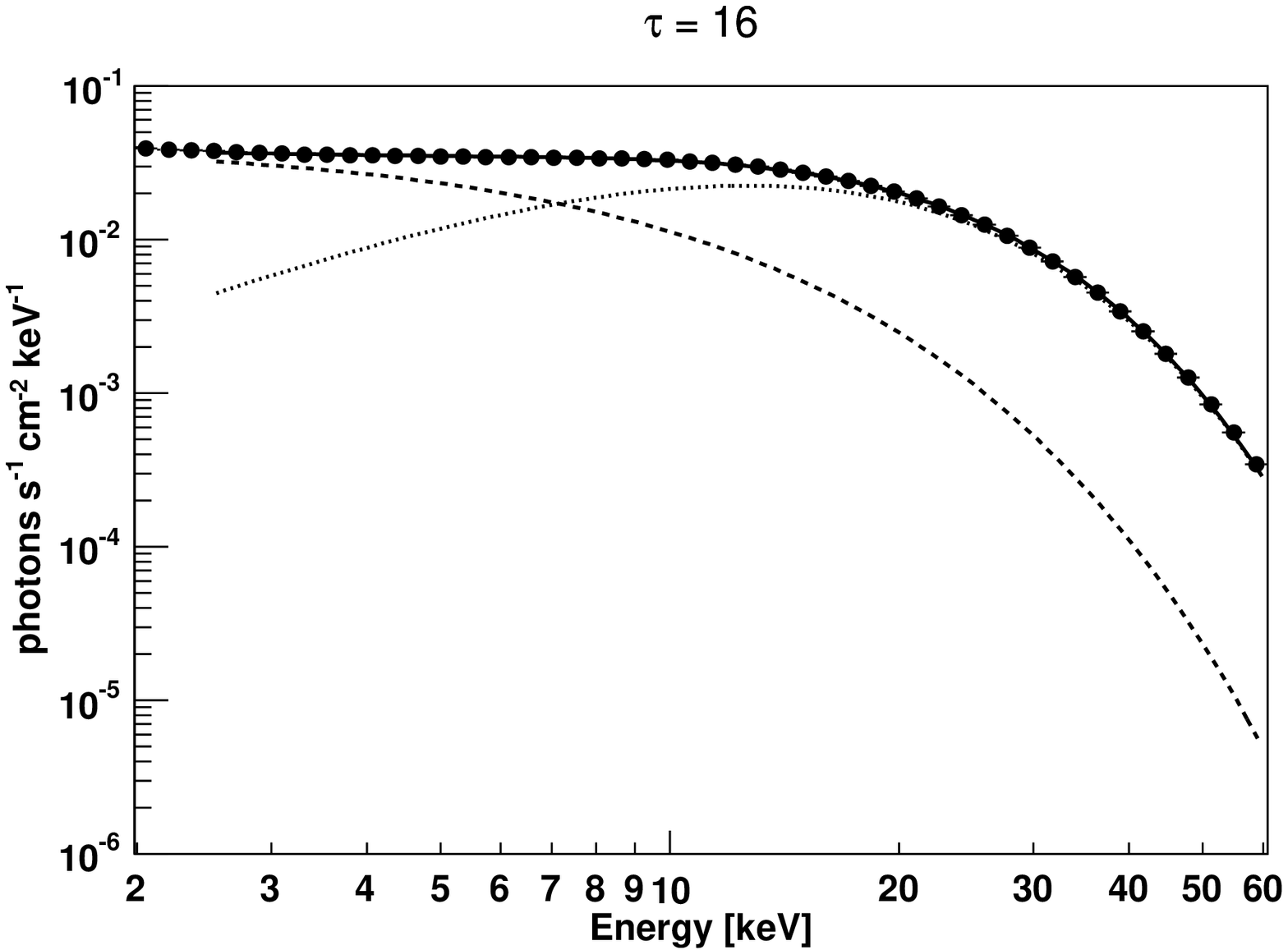}
\caption{Examples of simulation spectral models of thermal Comptonization. These models assume a common electron temperature $kT=6\ \mathrm{keV}$ and different radial optical thicknesses $\tau=1,\ 8,\ 16$ (as indicated at the top of each panel). Fitted NPEX models (solid), the negative (dashed) and the positive (dotted) power-law components are superposed. Statistical errors due to Monte Carlo sampling are sufficiently small compared with the filled circles showing the simulation data points.}
\label{fig:thermal_comptonization_model_large}
\end{center}
\end{figure}

The calculation was conducted for different Thomson thicknesses from
the cloud center to the cloud surface in a range between 1 and
20. These model calculations allow us to relate the phenomenological
parameters in the NPEX model to the physical characteristics of the
emitting system.  In
Figure~\ref{fig:thermal_comptonization_model_large}, we show spectra of thermal Comptonization for selected values, $\tau=1$, $8$ and $16$, of the optical depth. The
simulation results are fitted to the NPEX model
(Eq.~\ref{eq:model_npex}) in the energy range of 2.5--60 keV.  Though
we are now interested in the spectral shape only, the normalizations
of the spectra are determined by the required luminosity of the seed
photons of $L=1\times 10^{36}\ \mathrm{erg\ s^{-1}}$ in a range of
0.1--100 keV at the distance of 1.9 kpc.  The fitting results show
that thermal Comptonization spectra are perfectly characterized by
the NPEX model.


\begin{figure*}[tbp]
\begin{center}
\includegraphics[width=8.1cm]{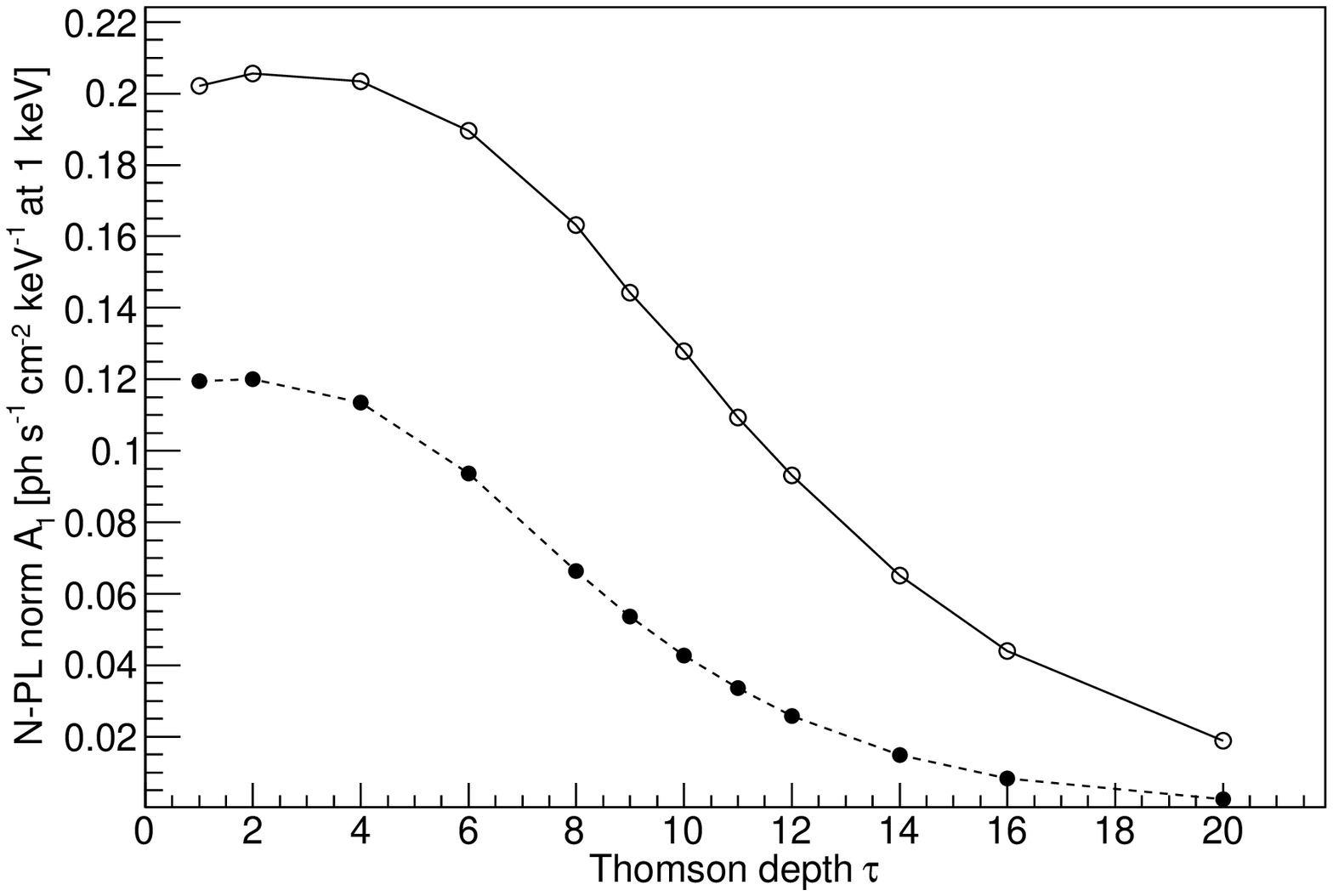}
\includegraphics[width=8.1cm]{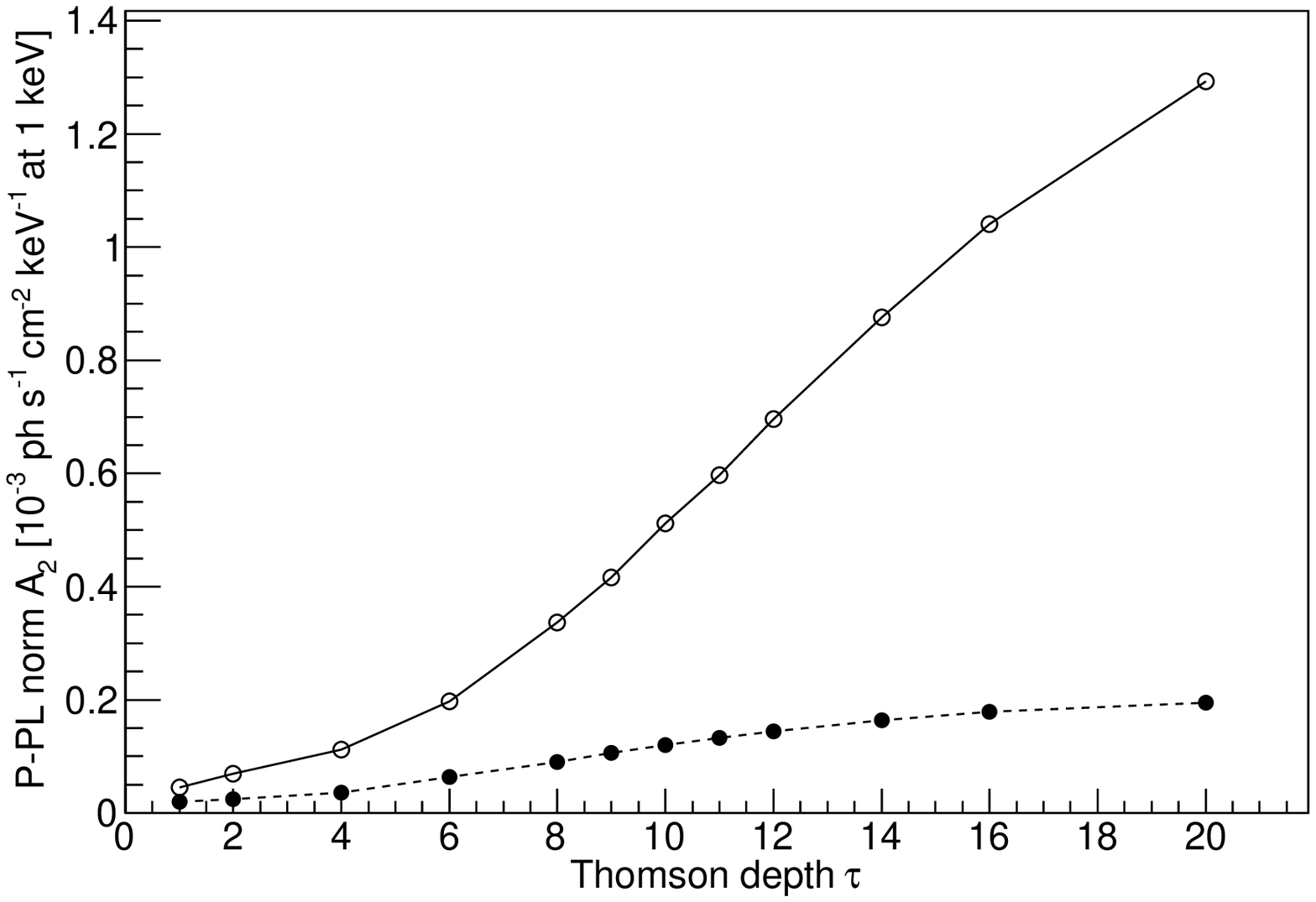}
\includegraphics[width=8.1cm]{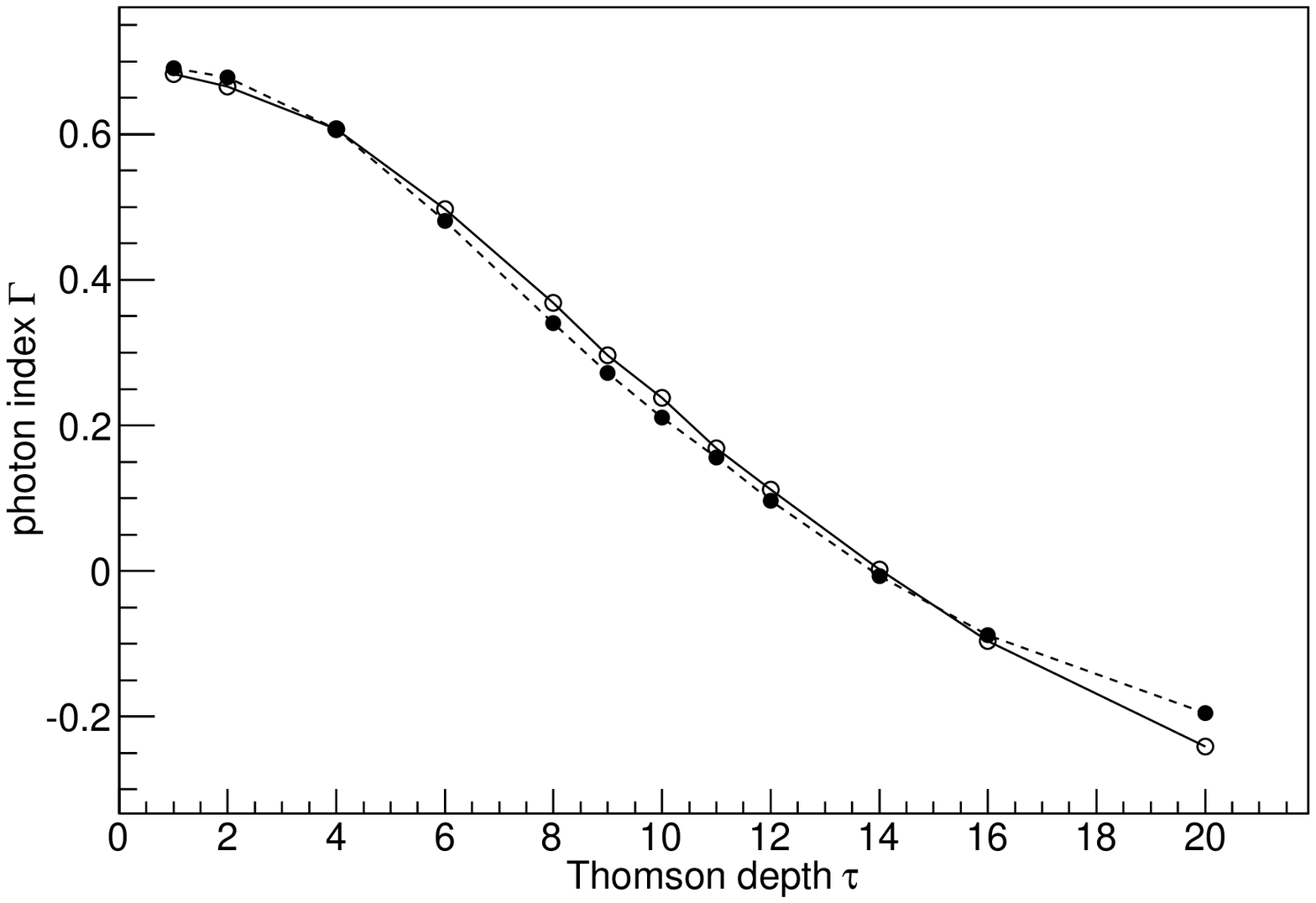}
\includegraphics[width=8.1cm]{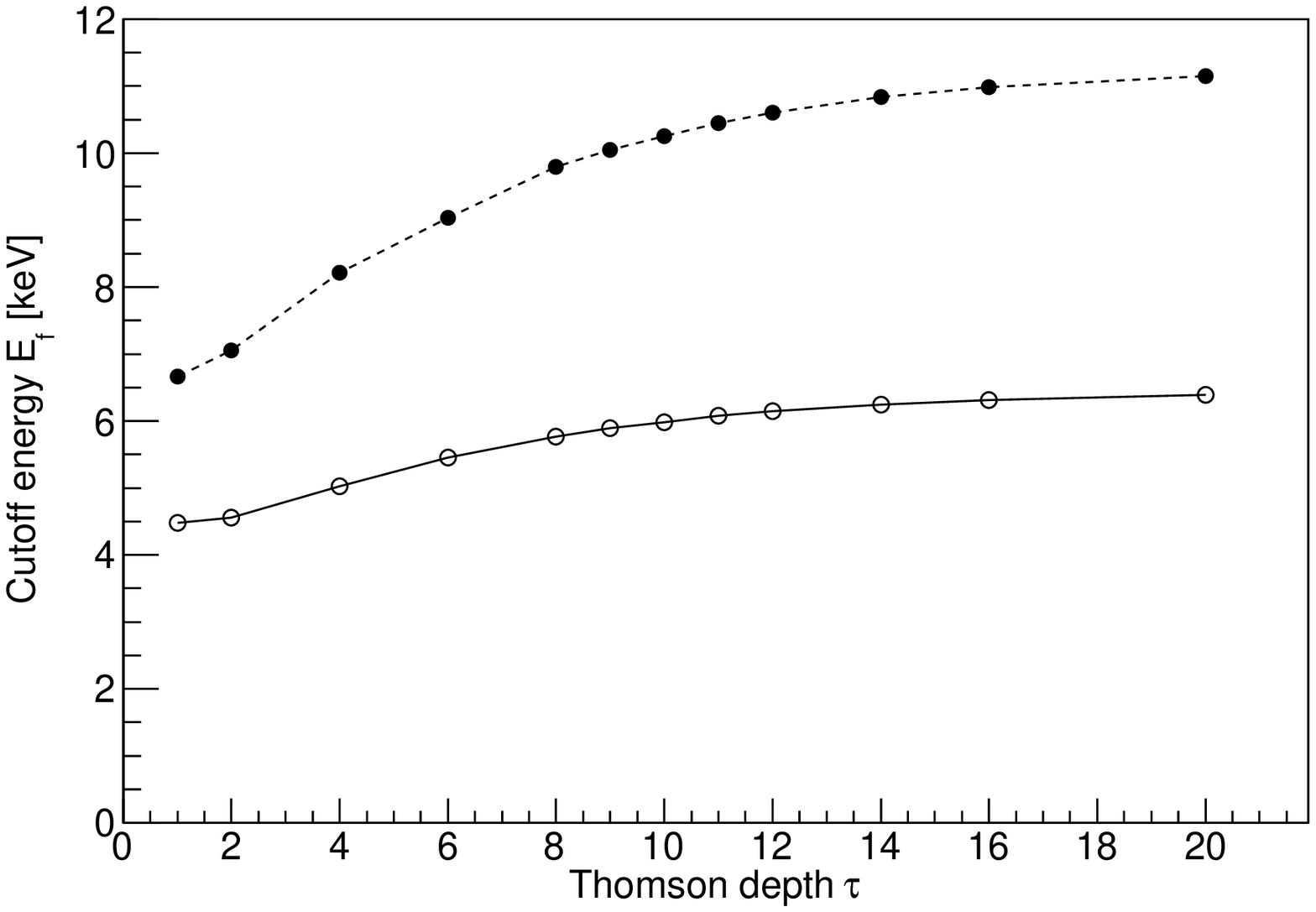}
\caption{Fitted NPEX model parameters, the negative power-law (N-PL) normalization $A_1$ (top-left), the positive power-law (P-PL) normalization $A_2$ (top-right), the photon index $\Gamma$ of the negative power law (bottom-left), and the cutoff energy $E_f$ (bottom-right), as functions of the optical thickness $\tau$ of the clouds. The solid line with open circles and the dashed line with filled circles are the functions for $kT=6$ keV and $kT=10$ keV, respectively.}
\label{fig:thermal_Comptonizaton_fit_parameters}
\end{center}
\end{figure*}

Figure~\ref{fig:thermal_Comptonizaton_fit_parameters} shows the fitted
model parameters as functions of the optical thickness $\tau$.
Importantly, the photon index of the negative power law becomes hard
with the optical thickness, being independent of the electron
temperature in the considered  range. Thus, the photon index $\Gamma$ can be
used as a good indicator of the optical thickness:
\begin{equation}\label{eq:tau_fit}
\tau=14-1.85\Gamma\,.
\end{equation}
The cutoff energy $E_f$ is comparable to the electron temperature, but
also changes with the optical thickness.  The normalizations of the two
power-law components have opposite dependence on the optical thickness,
 showing an approach to the saturation of the thermal
Comptonization. However, since these normalization coefficients scale
with the total flux and depend on the spectral slope, their absolute
values are not suitable for spectral characterization.

\begin{figure*}[tbp]
\begin{center}
\includegraphics[width=8.5cm]{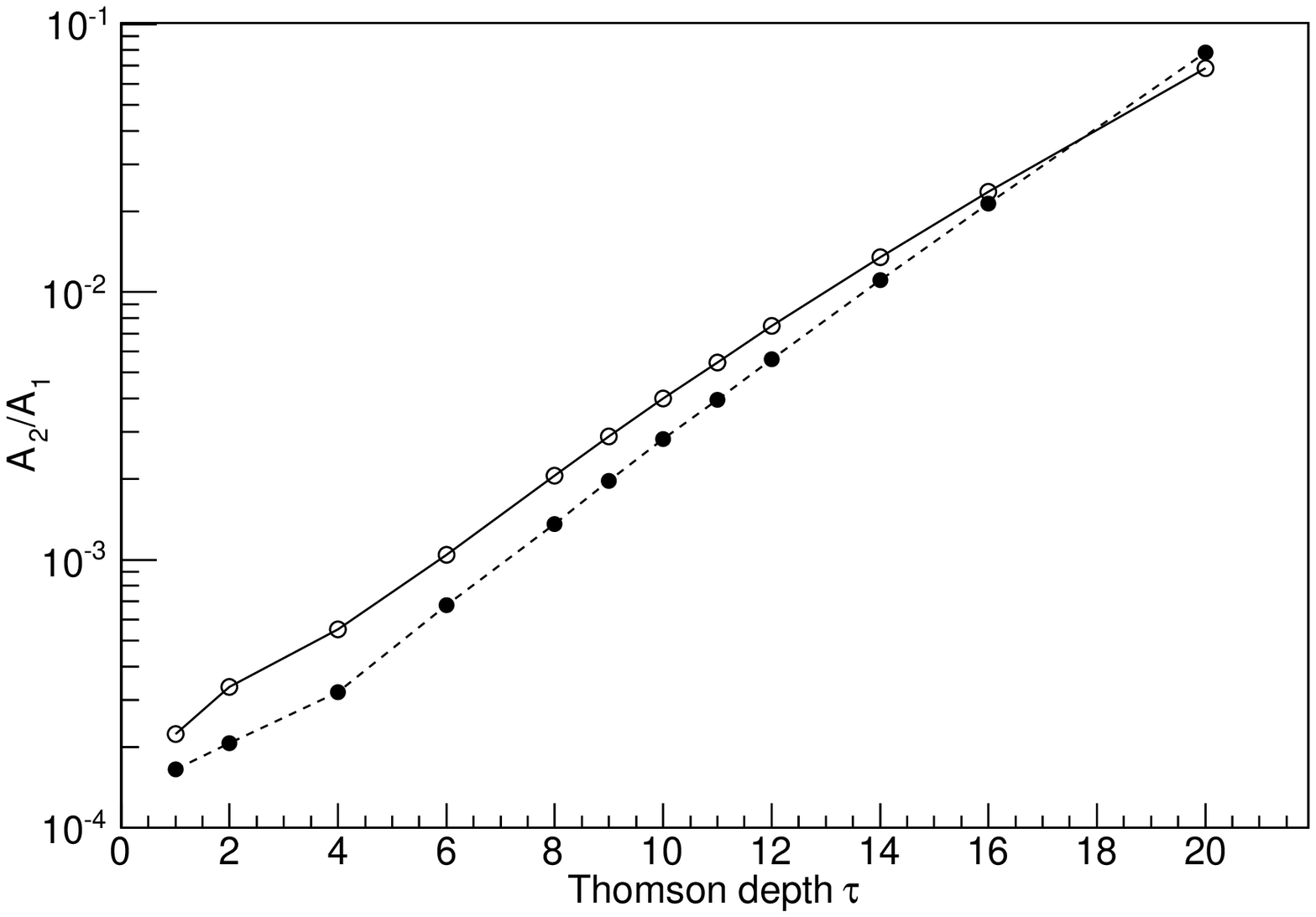}
\includegraphics[width=8.5cm]{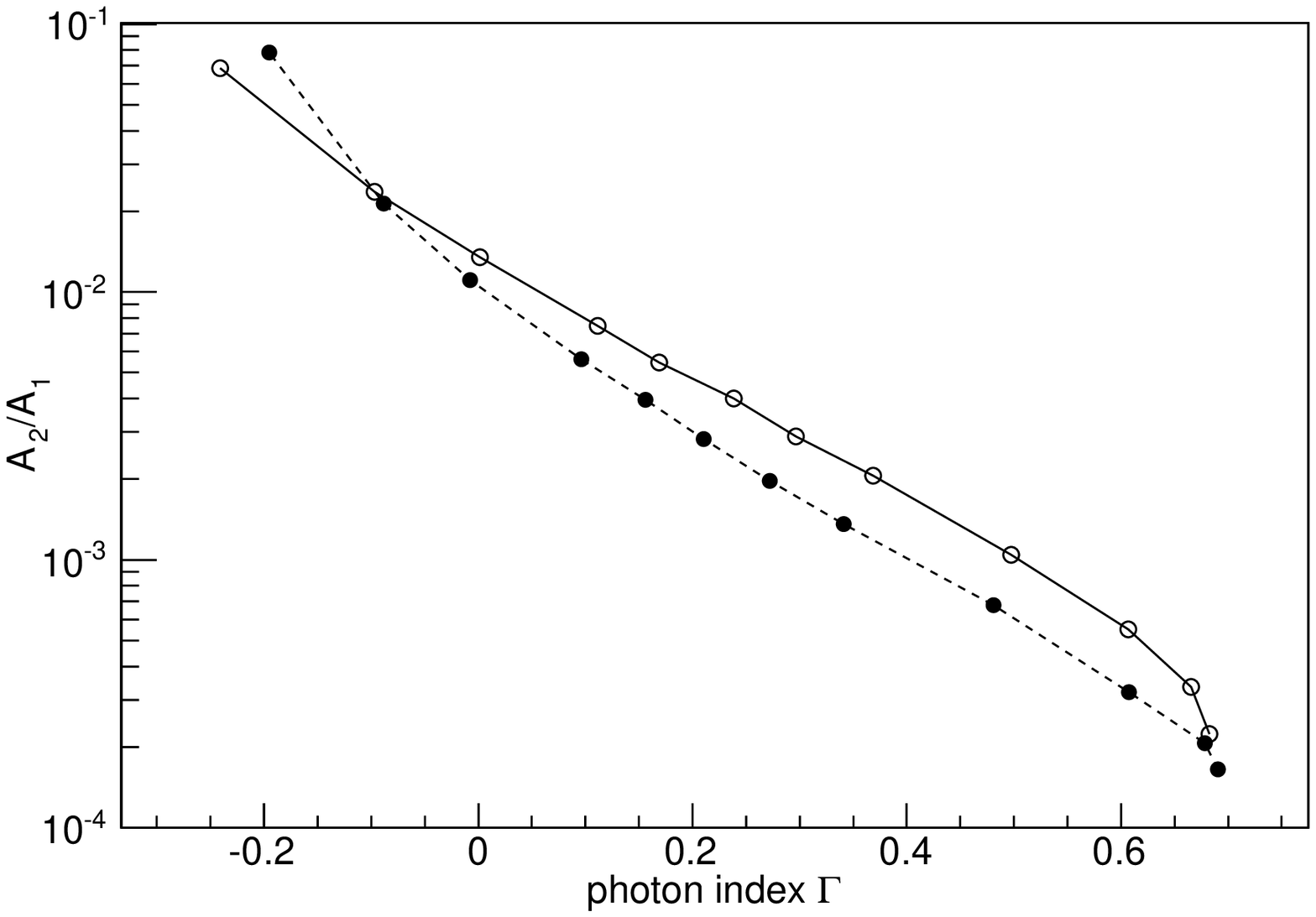}
\caption{Left: ratio $A_2/A_1$ between the normalizations of two power-law components of the NPEX model as a function of the optical thickness $\tau$. Right: relation between $A_2/A_1$ and the photon index $\Gamma$ of the negative power-law component. The solid line with open circles and the dashed line with filled circles are the functions for $kT=6$ keV and $kT=10$ keV, respectively.}
\label{fig:thermal_Comptonizaton_pn_ratio}
\end{center}
\end{figure*}

Since the spectral slope weakly depends on physical properties except for the optical thickness, we propose that the ratio $A_2/A_1$ can also be used to indicate the degree of Comptonization, or the optical thickness of the Comptonizing cloud.
To confirm it, we show the relation between $A_2/A_1$ and the optical thickness $\tau$ in the left panel of Figure~\ref{fig:thermal_Comptonizaton_pn_ratio}.
The ratio $A_2/A_1$ has strong dependence on the optical thickness with a relatively small difference due to the electron temperature, and moreover it can be regarded as a monotonically increasing function of the optical thickness.
In addition, it is helpful to see the relation between $A_2/A_1$ and the photon index $\Gamma$ of the negative power-law component, which is also a good indicator of $\tau$, as shown in the right panel of Figure~\ref{fig:thermal_Comptonizaton_pn_ratio}.
The normalization ratio $A_2/A_1$ significantly increases as the spectrum becomes hard.
If thermal Comptonization plays an important role in generating X-ray radiation in an astrophysical object that shows spectral variability, such a remarkable correlation between $A_2/A_1$ and the spectral index, both of which can be directly determined by observation, should appear.

\subsection{Comparison with Observations}

\begin{figure}[tbp]
\begin{center}
\includegraphics[width=8.5cm]{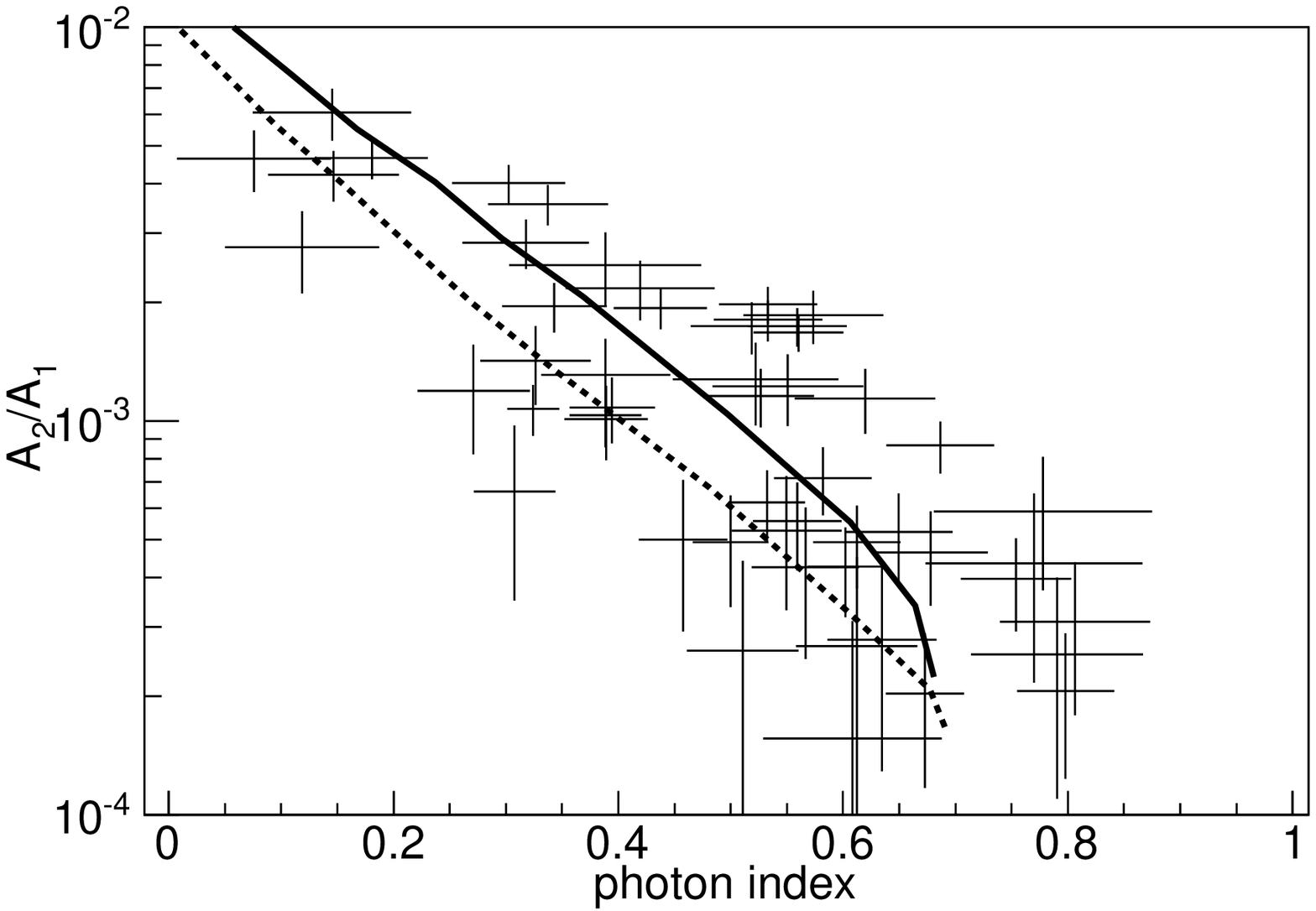}
\caption{Relation between $A_2/A_1$ and the photon index $\Gamma$ of
  the negative power-law component extracted from the \suzaku analysis
  of Vela X-1. Error bars indicate 1$\sigma$ uncertainties. The
  simulation models for $kT=6\ \mathrm{keV}$ (solid) and $kT=10\
  \mathrm{keV}$ (dashed) are also superposed.}
\label{fig:observation_velax1_pn_ratio}
\end{center}
\end{figure}

Based on the simulation results presented in the previous section, we conclude that thermal Comptonization model implies a
correlation between the power-laws' normalization ratio $A_2/A_1$ and the photon index $\Gamma$ of the NPEX model.  Since in
\citetalias{Odaka:2013} we reported time dependent spectral properties of \vela obtained with \suzaku (the entire observation
period of 145~ks was resolved over short periods of 2~ks), it is possible to test the inferred correlation against this set of
observational data. In Figure~\ref{fig:observation_velax1_pn_ratio},  we show  a comparison between the model prediction and the data.  It can be seen that the correlation  nicely agrees with the theoretical model shown
in the right panel of Figure~\ref{fig:thermal_Comptonizaton_pn_ratio}. We note that two exceptional points
which show a very soft index $\Gamma\sim 1.7$  at a low-state \citepalias[ID: 17 in][]{Odaka:2013} and a very hard index $\Gamma\sim -0.2$ at a
sharp flaring  \citepalias[ID: 48 in][]{Odaka:2013} were removed from the plot.  

While the positive power-law normalization $A_2$ is mathematically coupled with the photon index $\Gamma$, the
large change of $A_2/A_1$ ranging over two orders of magnitude is consistent with changing the optical depth of the Comptonizing cloud.  This result,
therefore, suggests that thermal Comptonization can play an important role in generating the X-ray spectrum from \vela.

\begin{figure}[tbp]
\begin{center}
\includegraphics[width=8.5cm]{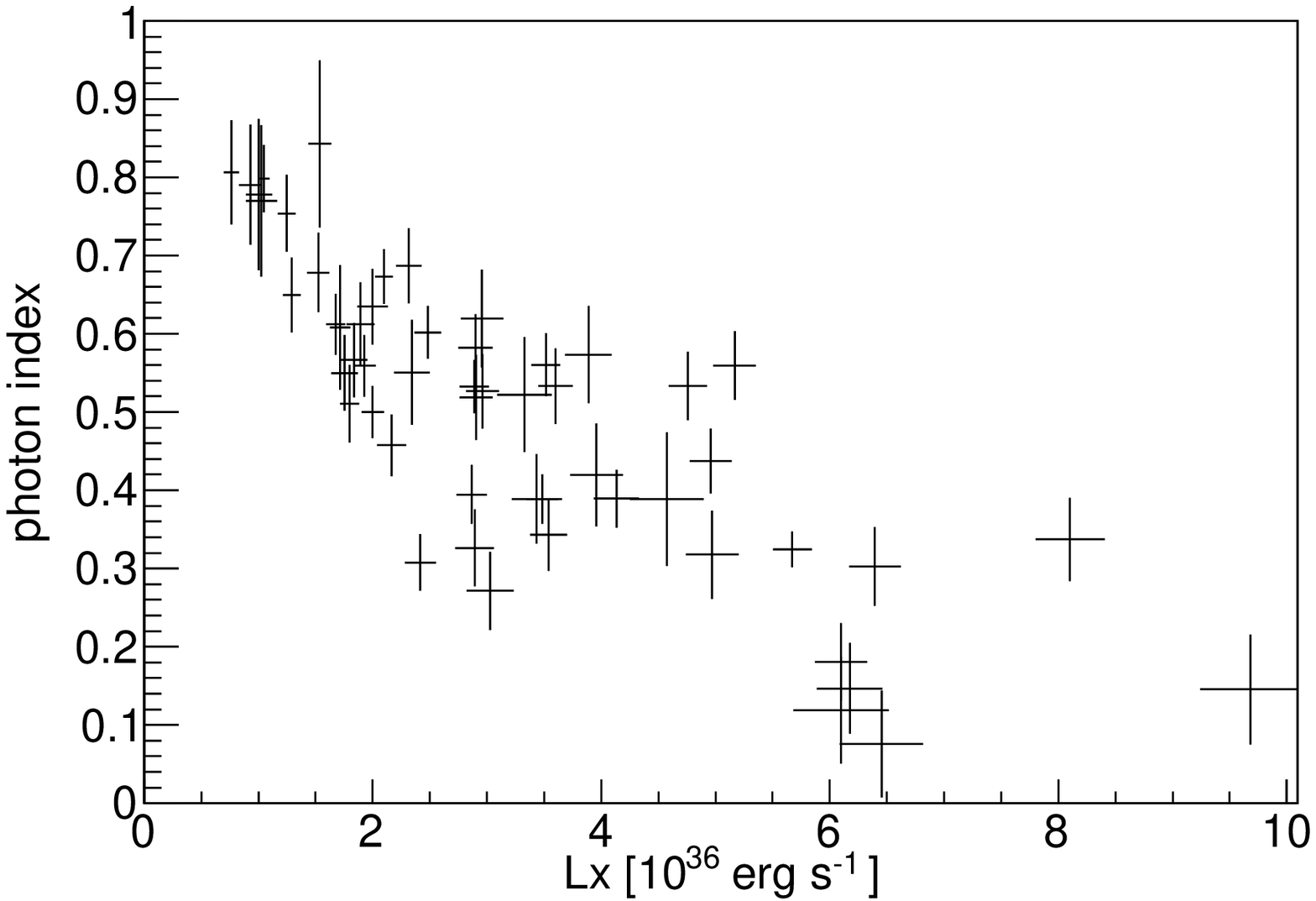}
\caption{Relation between the photon index $\Gamma$ and
the X-ray luminosity. This plot is taken from Fig.~19 in \citetalias{Odaka:2013}.}
\label{fig:observation_velax1_gamma_Lx}
\end{center}
\end{figure}

\begin{figure}[tbp]
\begin{center}
\includegraphics[width=8.5cm]{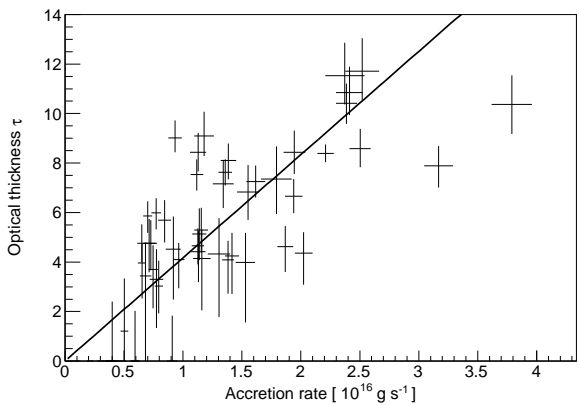}
\caption{Relation between the effective optical thickness $\tau$ and
the accretion rate $\dot{M}$. The solid line shows a linear
function described by Equation (\ref{eq:relation_tau_Mdot_fitted}).}
\label{fig:observation_velax1_tau_mdot}
\end{center}
\end{figure}

As shown in Figure~\ref{fig:observation_velax1_gamma_Lx}, the data suggest a
phenomenological correlation between the photon index and X-ray
luminosity, which can be converted into a relation between some key
parameters describing the conditions in the accretion flow.
Namely, the source luminosity is directly related to the mass accretion rate
$\dot{M}$ ($L\simeq GM_*\dot{M}/R_*$) and the photon index is linked to the production region opacity by
Equation~(\ref{eq:tau_fit}). Therefore, we obtain a relation between the accretion
rate $\dot{M}$ and the effective optical thickness $\tau$ of the
accretion plasma, as shown in
Figure~\ref{fig:observation_velax1_tau_mdot}.  Within the context of the NPEX model the accretion rate and
the optical thickness have a positive correlation.  A linear function
to describe the relation is
\begin{equation}\label{eq:relation_tau_Mdot_fitted}
\tau = 4.3\times \frac{\dot{M}}{10^{16}\ \mathrm{g\ s^{-1}}}.
\end{equation}
Although, the disagreement between the observational points and the
simple relation Equation~(\ref{eq:relation_tau_Mdot_fitted}) indicates a
possible impact of other physical processes, the general tendency of
the correlation of the optical depth and accretion rate supports the
scenario of Comptonized X-ray radiation.

\section{Thermal and Bulk Comptonization in Accretion Column Model}
\label{sec:accretion_column_model}

The pure thermal Comptonization model described in the previous
section lacks geometrical and dynamical structure of the column and
does not account for the impact of the magnetic field. Since these factors can
introduce differences to the simple picture considered above, a
physical model of the accretion column requires a more detailed
treatment.  In this section, we describe a self-consistent physical
model of an accretion column on a neutron star based on thermal and
bulk Comptonization.

Before proceeding to a more detailed calculation, however, it is possible to obtain first order approximation to the structure of the column
based on the results obtained in Section 3.  Namely,  assuming  a simple geometry of the
accretion column with a radius of $r_0$ in which the number density $n(z)$ and the bulk velocity $v(z)$ of the plasma depend only on the
coordinate $z$, the  accretion rate can be written as
\begin{equation}\label{eq:mdot_column} \dot{M} = \pi r_0{}^2 n(z)
m_\mathrm{p} |v(z)|.
\end{equation} 
The optical thickness in the horizontal direction ($r$-direction) from
the center to the wall surface is
\begin{equation}\label{eq:tau_column} \tau=r_0n(z)\sigma_\perp.
\end{equation} 
Substituting Equations~(\ref{eq:mdot_column}) and
(\ref{eq:tau_column}) to Equation~(\ref{eq:relation_tau_Mdot_fitted}), we
obtain the column radius as
\begin{equation} r_0 = \frac{\sigma_\mathrm{T}}{\pi
m_\mathrm{p}}\frac{\dot{M}}{\tau}\frac{1}{v} = 1.0\times 10^5
\left(\frac{0.1c}{v}\right) \ \mathrm{cm}.
\end{equation} 
Here, we assumed $\sigma_\perp=\sigma_\mathrm{T}$.  If
the velocity of the decelerated accretion flow is $0.1c$ near the
surface, the radius of the accretion column in Vela X-1 is $\sim 1$
km.  This value is natural by comparison with the neutron star radius
$R_*\sim 10$ km, which reinforces the hypothesis of the emission
mechanism due to Comptonization.

\subsection{Simulation of Accretion Column}\label{subsec:simulation_column}

Based on the dynamical structure described above, we performed Monte
Carlo simulations of Comptonized X-ray radiations.  The calculation
methods are given in Appendix \S\ref{sec:calculation}.  The geometry
including the distributions of the number density and the velocity is
completely specified by three parameters: the mass accretion rate
$\dot{M}$, the column radius $r_0$, and the height of the sonic point
$z_\mathrm{sp}$.  The accretion rate is determined by the luminosity
$L_X$ if the mass $M_*=1.9M_\odot$ \citep{Quaintrell:2003}, and radius
$R_*=10^{6}\ \mathrm{cm}$ of the neutron star are fixed.  The sonic
point $z_\mathrm{sp}$ is represented as $z_\mathrm{sp}=\xi r_0$, where
we introduce a dimensionless scaling parameter $\xi$. 
Equation~\ref{eq:sonic_point} shows that according to
the analytical solution obtained by \citet{Basko:1976} and \citet{Becker:1998}, the value of this
parameter should depend on the ratio of the energy-averaged cross sections that
determine interaction rates for photons propagating parallel and
perpendicular to the magnetic field, and is expected to be almost constant {\bf if the energy spectrum of the radiation does not largely change.}

According to the velocity profile
(Equation~\ref{eq:column_velocity_profile}), the velocity reaches zero at
the surface of neutron star, and therefore the number density becomes
very large. It implies the existence of a ``photosphere'', i.e.\ a height
below which the density is sufficiently large to be optically thick.
Since photon absorption occurs via free-free absorption in this
regime, the position of the photosphere can be estimated as a point at
which horizontal optical thickness of free-free absorption becomes
unity \citep{Becker:2007}.  This relation is represented as
\begin{equation}
\alpha_R^\mathrm{ff} r_0 = 1
\end{equation}
where $\alpha_R^\mathrm{ff}$ denotes the Rosseland-mean of free-free absorption coefficient, and can be represented in cgs units as \citep{Rybicki:radiative_process}
\begin{equation}
\alpha_R^\mathrm{ff} = 1.7\times 10^{-25}\ T^{-7/2}n^2 \ \mathrm{cm^{-1}}.
\end{equation}
We regard the position $z_\mathrm{b}$ given by this relation as the
bottom of the accretion column, and this surface at $z=z_\mathrm{b}$
can be treated as a blackbody emitter.

We have used the Monte Carlo approach to calculate the spectrum of the
photons escaping from the accretion column, based on the source
function of seed photons \citep[see Appendix A of][for the description
of the numerical approach]{Odaka:2011}.  In the accretion column the
seed photons are provided by thermal bremsstrahlung from the plasma,
cyclotron radiation of electrons in the strong magnetic field, and
blackbody radiation from the bottom of the column.  In this paper we
compute only the Comptonization of the thermal bremsstrahlung photons,
other contributions will be considered elsewhere. However, we note that
in the case of {\bf high luminosity} sources, bremsstrahlung provides the dominant
donor of seed photons among the three processes \citep{Becker:2007}.

For thermal bremsstrahlung, the source function, or a photon
production rate, is given by
Equation~(\ref{eq:source_function_freefree}).  As the emissivity is
proportional to the density squared, $n^2$, the seed photons
concentrate near the bottom of the column, where the density is large
due to the deceleration of the accretion flow.  For the sake of
simplicity, we assume that the seed photons are isotropic, though the
strong magnetic field can significantly affect the direction
distribution of the free-free emission.  A beaming effect due to the
bulk motions is negligible since most of the seed photons come from
the bottom region of the column, where the falling flow is
sufficiently decelerated.

\begin{figure}[tbp]
\begin{center}
\includegraphics[width=8.5cm]{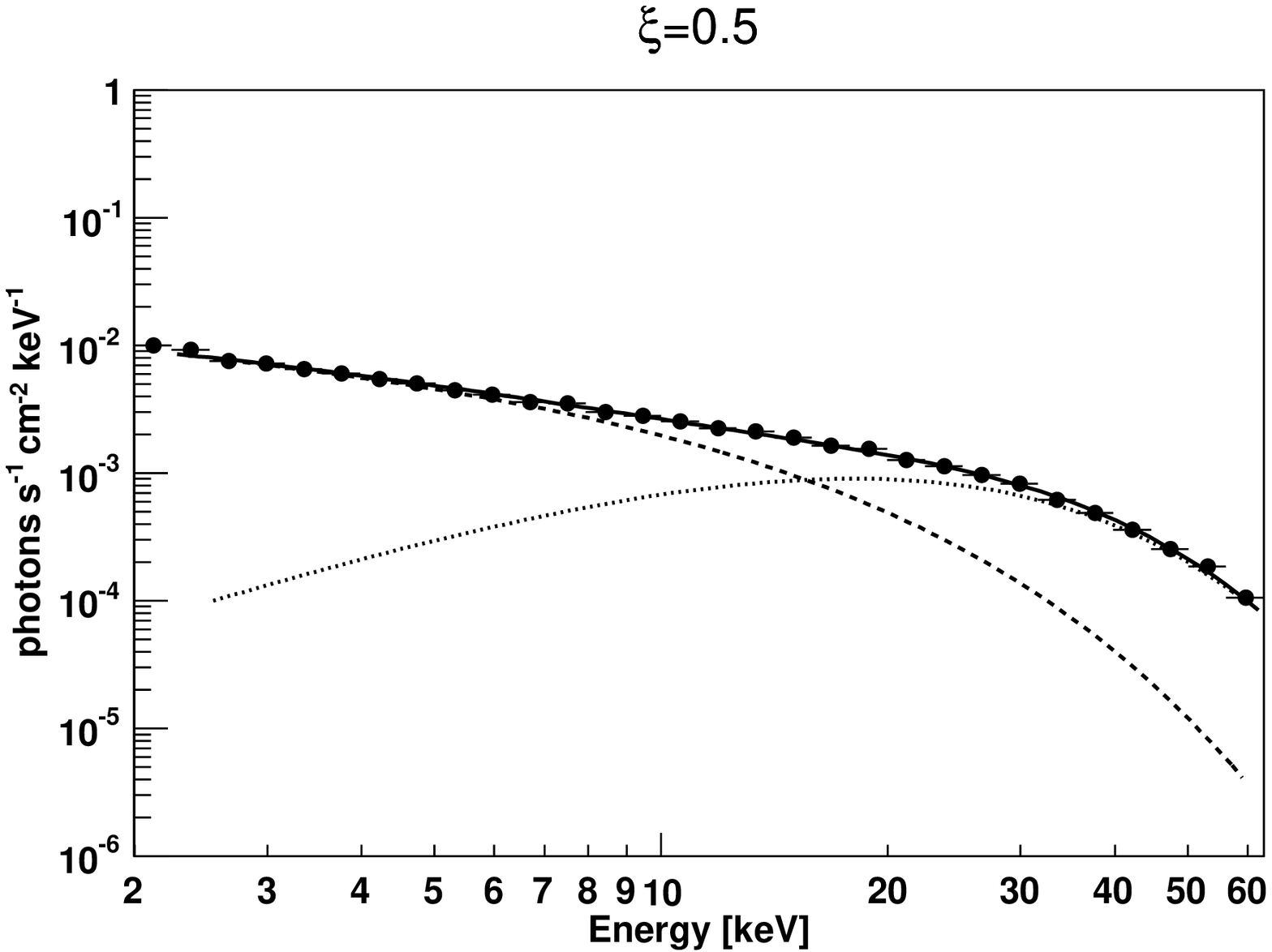}\\
\vspace{12pt}
\includegraphics[width=8.5cm]{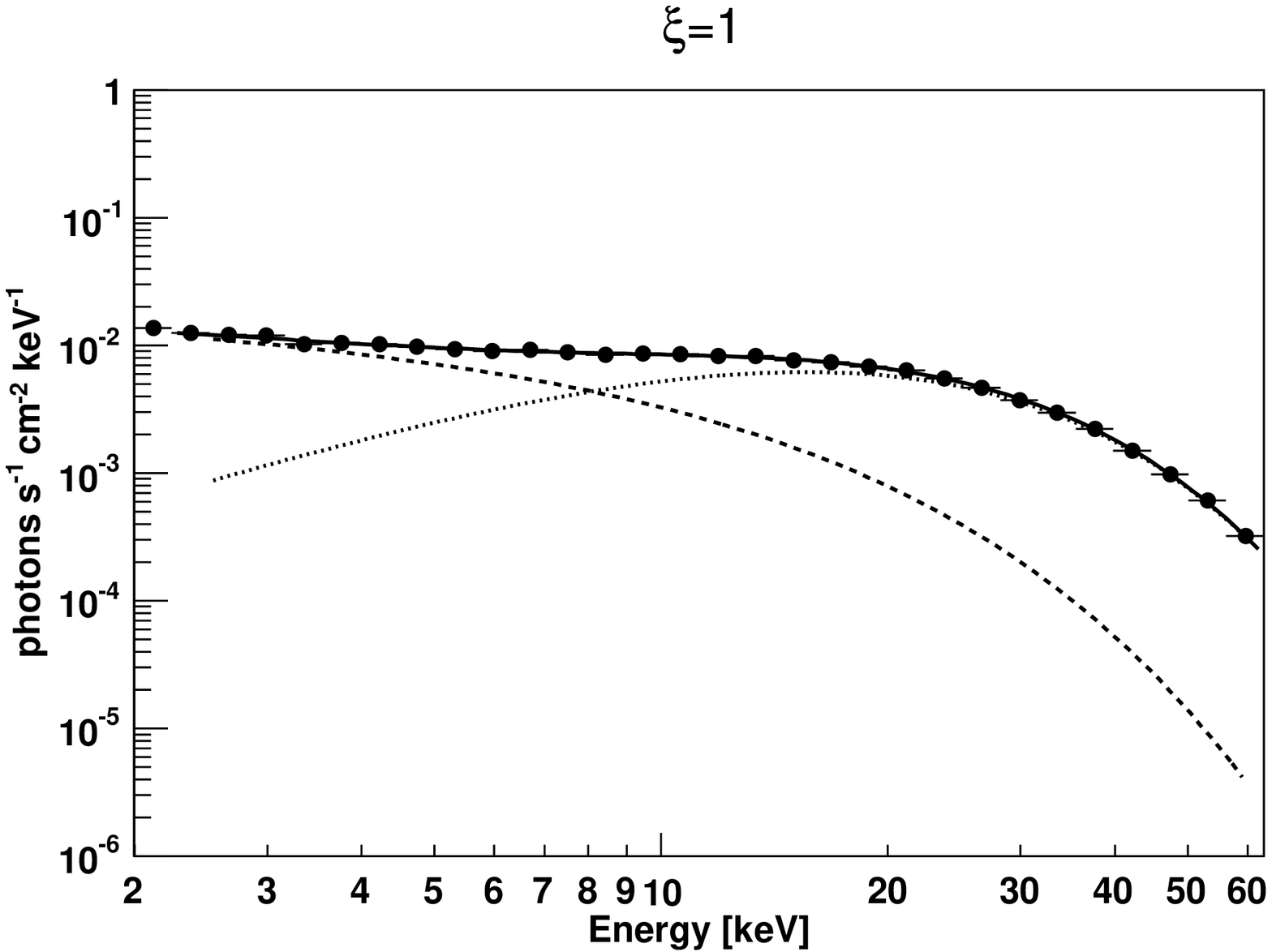}\\
\vspace{12pt}
\includegraphics[width=8.5cm]{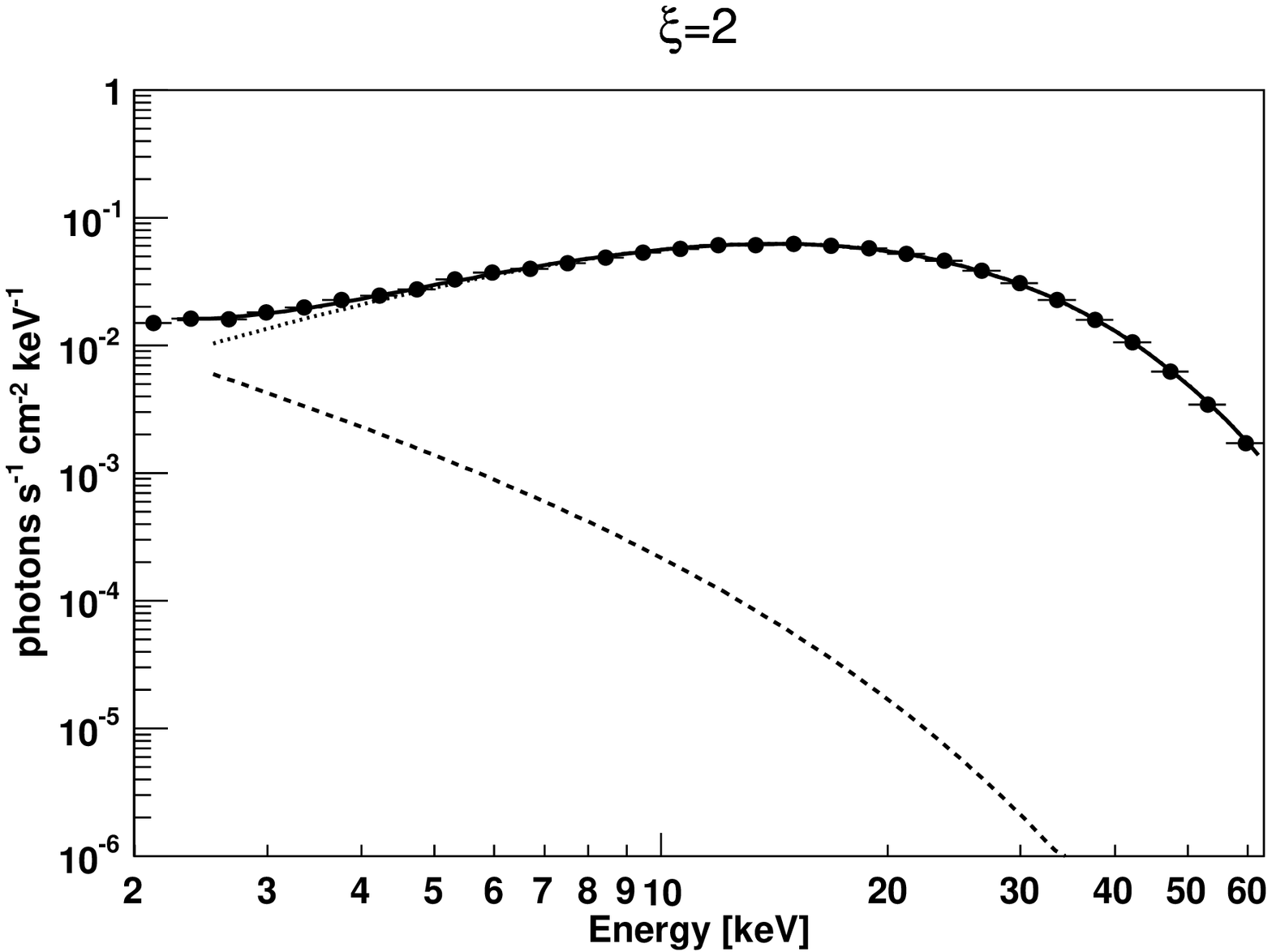}
\caption{Simulation spectra of thermal and bulk Comptonization from the accretion column. The simulation parameters are tabulated in Table \ref{table:simulation_parameter_column}. The value of the $\xi=z_\mathrm{sp}/r_0$ is indicated at the top of each panel. Fitted NPEX model (solid), the negative (dashed) and the positive (dotted) power-law components are superposed. The model parameters obtained by the fit are tabulated in Table~\ref{table:result_column_model_B0_kT6keV}. Statistical errors due to Monte Carlo sampling are sufficiently small compared with the filled circles showing the simulation data points.}
\label{fig:column_model_B0_kT6keV}
\end{center}
\end{figure}

The simulation parameters are listed in
Table~\ref{table:simulation_parameter_column}.  In these simulations,
we assumed a column radius $r_0=4\times 10^{4}$ cm; plasma
temperature $kT=6$ keV; and three values of $\xi=z_\mathrm{sp}/r_0=0.5$,
1.0, 2.0.  The spectra obtained by the simulations and fitted
parameters of the NPEX model are shown in
Figure~\ref{fig:column_model_B0_kT6keV} and Table
\ref{table:result_column_model_B0_kT6keV}, respectively.  The spectral
shape strongly depends on $\xi$ because this parameter controls the
density of the decelerated plasma.  For $\xi=2.0$, the spectrum is
nearly saturated, being dominated by the Wien spectrum \citep[for detail, see, e.g.,][\S7.7]{Rybicki:radiative_process}.
Note that for this case, the fitted values regarding the negative power-law term, i.e.\ $\Gamma$ and $A_1$, are no longer good indicators of the degree of Comptonization, since the fitted positive power-law term of the NPEX function completely dominates over the negative power-law term (see Table~\ref{table:result_column_model_B0_kT6keV} and the right panel of Figure~\ref{fig:column_model_B0_kT6keV}). In other words, characterization using the NPEX function fails in this regime.

\begin{table}[htdp]
\caption{Simulation parameters of the accretion column}
\begin{center}
\begin{tabular}{cc}
\hline\hline
parameter & value \\
\hline
Luminosity $L_X$ (0.1--100 keV) & $3.0\times 10^{36}\ \mathrm{erg\ s^{-1}}$ \\
Neutron star mass $M_*$ & $1.9M_\odot$ \\
Neutron star radius $R_*$ & $10^6\ \mathrm{cm}$ \\
Column radius $r_0$ & $4.0\times 10^{4}\ \mathrm{cm}$ \\
Electron temperature $kT$ & 6 keV \\
$\xi=z_\mathrm{sp}/r_0$ & 0.5, 1.0, 2.0\\
\hline
\end{tabular}
\end{center}
\label{table:simulation_parameter_column}
\end{table}%

\begin{table*}[htdp]
\caption{Fitted values of the simulated spectra to the NPEX model}
\begin{center}
\begin{tabular}{lccccccc}
\hline\hline
$\xi$ & $L_0$ & $A_1$ & $\Gamma$ & $A_2$ & $E_f$ & $L_X$ & $A_2/A_1$ \\
 & $\mathrm{erg\ s^{-1}}$ & $\mathrm{ph\ s^{-1}cm^{-2}keV^{-1}}$ & & $\mathrm{ph\ s^{-1}cm^{-2}keV^{-1}}$ & keV & $\mathrm{erg\ s^{-1}}$ & \\
\hline
0.5 & $8.77\times 10^{35}$ & $1.5\times 10^{-2}$ & $0.41$ & $2.1\times 10^{-5}$ & 9.0 & $8.9\times 10^{35}$ & $1.4\times 10^{-3}$ \\
1.0 & $2.39\times 10^{36}$ & $1.9\times 10^{-2}$ & $0.20$ & $1.9\times 10^{-4}$ & 7.8 & $3.4\times 10^{36}$ & $1.0\times 10^{-2}$ \\
2.0$^\mathrm{a}$ & $4.44\times 10^{36}$& $4.0\times 10^{-2}$ & $1.65$ & $2.3\times 10^{-3}$ & 7.1 & $2.4\times 10^{37}$ & $5.8\times 10^{-2}$ \\
\hline
\end{tabular}
\end{center}
\begin{flushleft} {\scriptsize a) For this case, the characterization by the NPEX function fails. (See text in detail.)}
\end{flushleft}
\label{table:result_column_model_B0_kT6keV}
\end{table*}%

\subsection{Magnetic Field Effects}\label{subsec:magnetic_filed_effect}

The strong magnetic field of $B\sim 10^{12}$ G close to the neutron
star surface largely affects the interactions between photons and
electrons.  Namely, in such a strong field, electrons are allowed to
move only along the field lines, while the perpendicular motion is
quantized to Landau states.  The imposed kinematic restrictions
imply a significant change of the interaction cross-section, which
gets reduced, particularly for photons that propagate parallel to the
field.  In addition, transitions between discrete energy levels
corresponding to the ``perpendicular'' motion of electrons (Landau levels),
results in formation of the so-called cyclotron emission and absorption.

Although the physical processes of photons in a strong magnetic field
are very complex \citep[e.g.][]{Ventura:1979, Kirk:1980, Arons:1987},
we rely  here on a simple treatment of the photon propagation in the
magnetic field, which is based on our Comptonization model described in Appendix
\S\ref{sec:calculation}.  Since the reduction of the scattering cross
sections is the most significant effect of the magnetic field on the
photon propagation, the total cross sections for photon interactions are modified
in our treatment.  Such a strong magnetic field leads to a birefringent nature of the photon propagation.
Thus, we distinguish  scattering of two linearly polarized normal modes, ordinary and extraordinary.
The ordinary mode is linearly polarized with electric field vector that is placed in the plane formed by the
photon propagation direction and the magnetic field; while the extraordinary mode has electric
field vector perpendicular to this plane.  Approximate total
cross sections of the ordinary (1) and extraordinary (2) mode are
given by
\begin{gather}
\sigma^{(1)}(E) = \sigma_\mathrm{T} [\sin^2\theta + f(E)\cos^2\theta],\label{eq:cross_section_1} \\
\sigma^{(2)}(E) = \sigma_\mathrm{T} f(E),\label{eq:cross_section_2}
\end{gather}
respectively, where $\sigma_\mathrm{T}$ is the Thomson cross section, $\theta$ is the angle between the photon direction and the magnetic field, and the $f(E)$ is given by
\begin{equation}
f(E) = \left\{
\begin{array}{cc}
\left(\dfrac{E}{E_c}\right)^2 & (E\le E_c) \\
1 & (E\ge E_c) \\
\end{array}
\right. ,
\end{equation}
where $E_c$ denotes the cyclotron energy \citep{Arons:1987}.
Since cyclotron resonance scattering is effective only near the resonance energy $E_c$, we do not include this process in the total cross section of the scattering, though the cross section of the resonance effect is significant at $E=E_c$.

Although, the presence of the magnetic field changes the electron distribution and differential cross sections of the scattering processes, the other parts of the calculation, which include the sampling of the target electron and the determination of the scattered photon in the rest frame of the target, remain unmodified for simplicity.
Consequently, including the magnetic filed in our model of Comptonization (Appendix~\S\ref{sec:calculation}) only affects the determination of the next interaction point in terms of the Monte Carlo simulations.

\begin{figure}[tbp]
\begin{center}
\includegraphics[width=8.5cm]{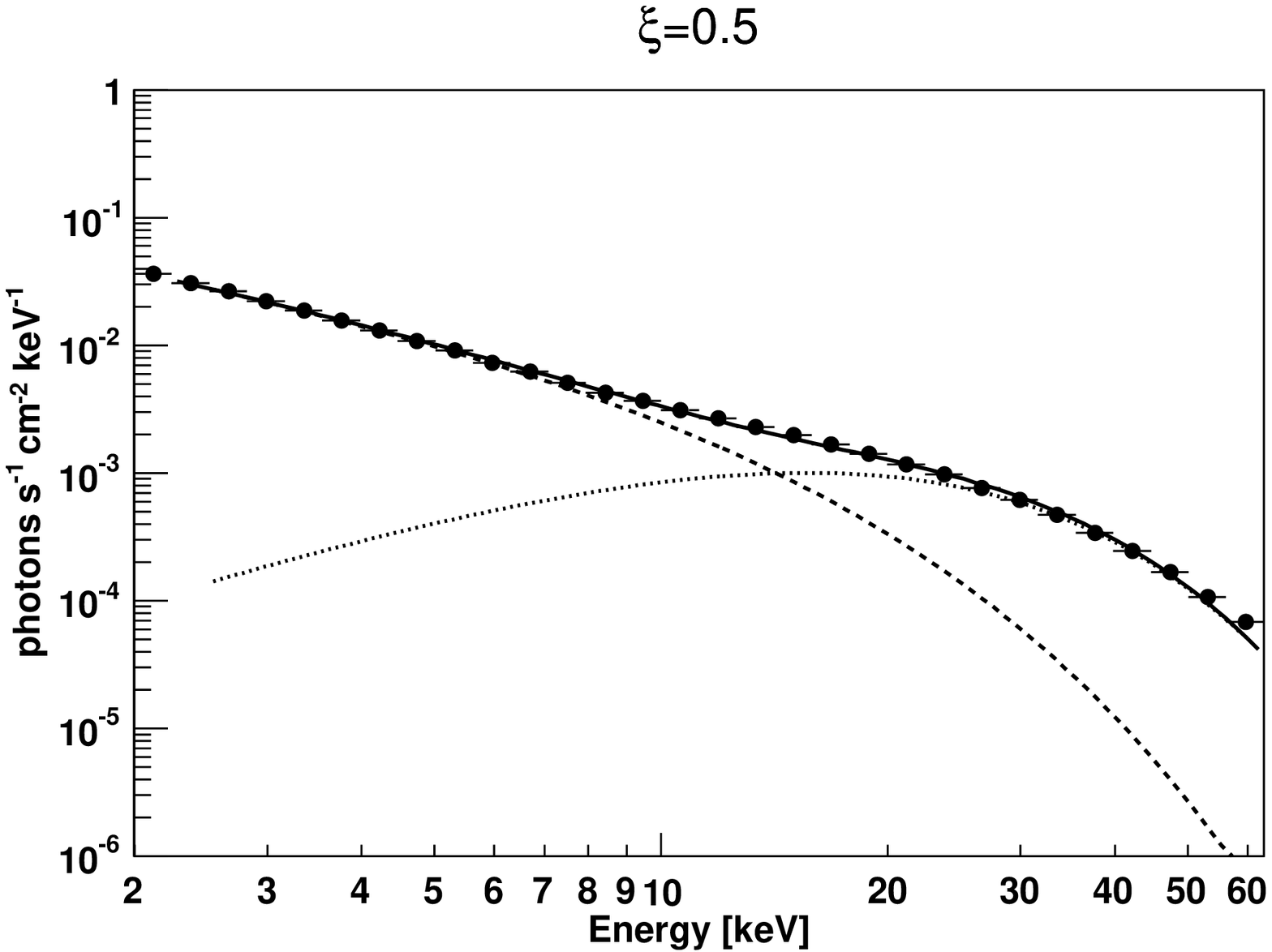}\\
\vspace{12pt}
\includegraphics[width=8.5cm]{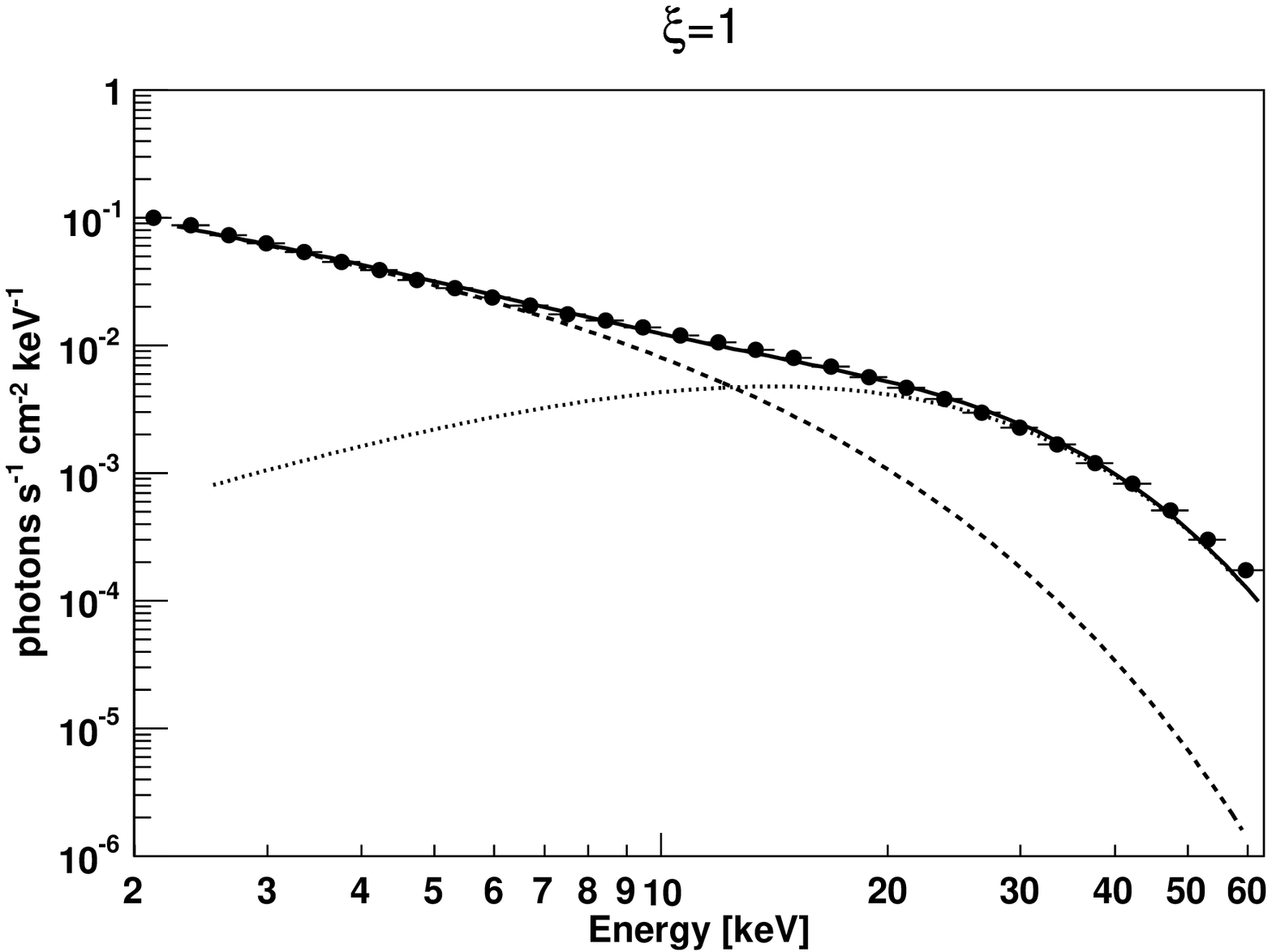}\\
\vspace{12pt}
\includegraphics[width=8.5cm]{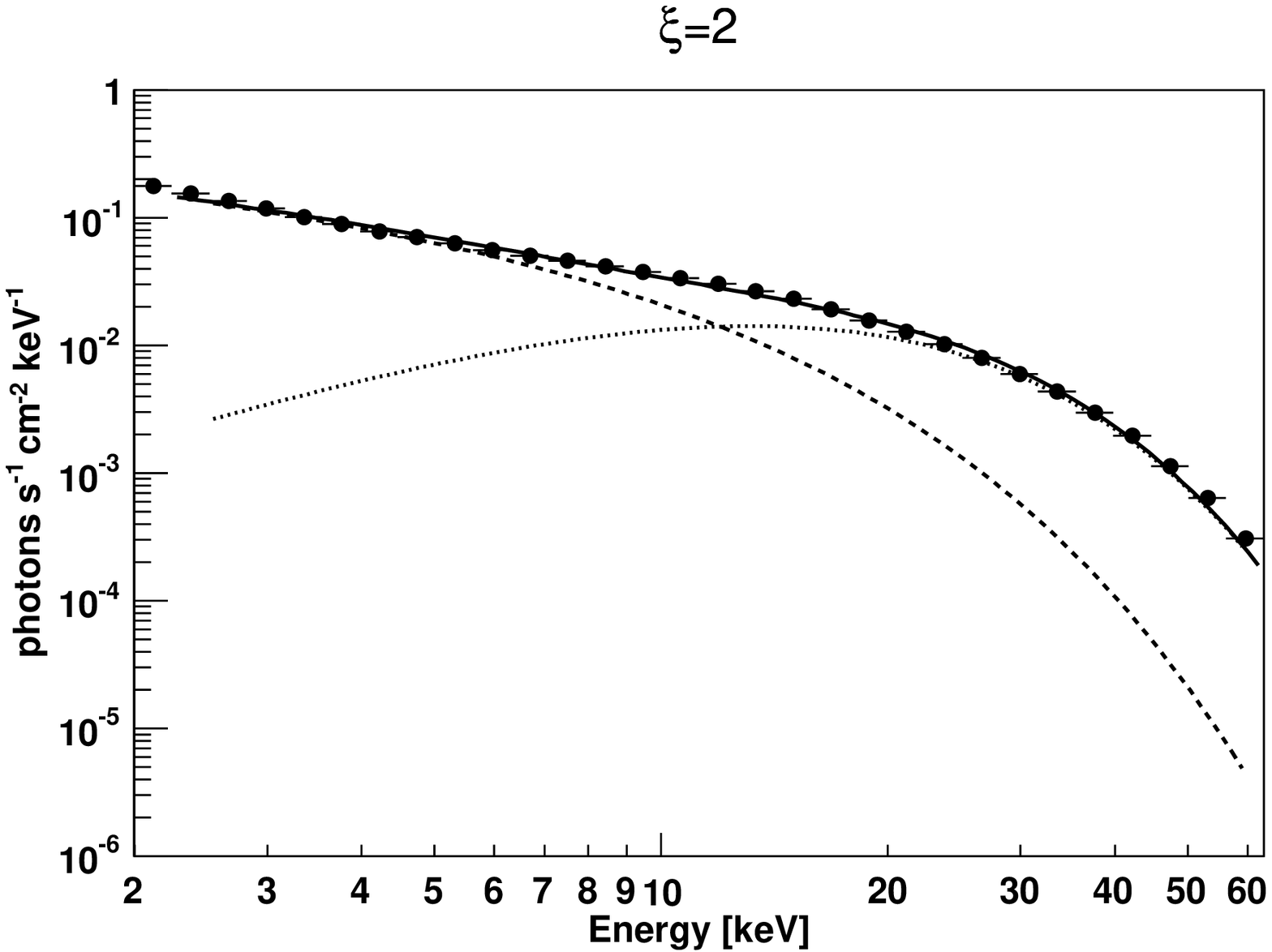}
\caption{Same as Figure~\ref{fig:column_model_B0_kT6keV} but for accretion columns that have a strong magnetic field of $B=2\times 10^{12}$ G. The model parameters obtained by the fit are tabulated in Table~\ref{table:result_column_model_B2_kT6keV}.}
\label{fig:column_model_B2_kT6keV}
\end{center}
\end{figure}

Simulation results based on this model are shown in Figure~\ref{fig:column_model_B2_kT6keV} and Table \ref{table:result_column_model_B2_kT6keV}.
The conditions of the accretion column are completely identical to the simulations demonstrated in the previous subsection (\S\ref{subsec:simulation_column}) except that these simulations account for the presence of magnetic field with $B=2\times 10^{12}$ G.
All the three spectra qualitatively agree with spectral features of accreting X-ray pulsars, which are represented by the NPEX model. 
The effect of the reduced cross sections is obvious from the mildly Comptonized spectrum even for $\xi=2.0$.
Since Comptonization becomes significant with $\xi$, the photon index $\Gamma$ and the normalization ratio $A_2/A_1$ increase with $\xi$.

\begin{table*}[htdp]
\caption{Fitted values of the simulated spectra with a magnetic field to the NPEX model}
\begin{center}
\begin{tabular}{lccccccc}
\hline\hline
$\xi$ & $L_0$ & $A_1$ & $\Gamma$ & $A_2$ & $E_f$ & $L_X$ & $A_2/A_1$ \\
 & $\mathrm{erg\ s^{-1}}$ & $\mathrm{ph\ s^{-1}cm^{-2}keV^{-1}}$ & & $\mathrm{ph\ s^{-1}cm^{-2}keV^{-1}}$ & keV & $\mathrm{erg\ s^{-1}}$ & \\
\hline
0.5 & $8.77\times 10^{35}$ & $1.0\times 10^{-1}$ & $1.05$ & $3.1\times 10^{-5}$ & 7.8 & $9.7\times 10^{35}$ & $3.1\times 10^{-4}$ \\
1.0 & $2.39\times 10^{36}$ & $2.4\times 10^{-1}$ & $0.85$ & $1.8\times 10^{-4}$ & 7.0 & $3.2\times 10^{36}$ & $7.5\times 10^{-4}$ \\
2.0 & $4.44\times 10^{36}$& $3.1\times 10^{-1}$ & $0.51$ & $6.1\times 10^{-4}$ & 6.6 & $7.8\times 10^{36}$ & $2.0\times 10^{-3}$ \\
\hline
\end{tabular}
\end{center}
\label{table:result_column_model_B2_kT6keV}
\end{table*}%

\section{Discussion}\label{sec:discussion}

Our radiation model of thermal and bulk Comptonization in an accretion column with a strong magnetic field well reproduces the broad-band spectral
characteristics observed from accreting neutron stars.  Based on the model, we find self-consistent
solutions that agree well with the observed spectra of \vela. We study solutions where the accretion plasma itself provides the seed
photons through free-free emission.  Thus, the simulation model has only three parameters: the X-ray luminosity $L_\mathrm{obs}$
obtained by observation, the radius of the accretion column $r_0$, and the ratio of the the sonic-point height to the column radius
$\xi=z_\mathrm{sp}/r_0$.  While $r_0$ and $\xi$ determine the geometry of the column including the $z$-profiles of the velocity and
the density, the observed luminosity $L_\mathrm{obs}$ determines the accretion rate $\dot{M}$, i.e.\ the normalization of the density
distribution.  Once these three parameters fix the simulation geometry, the distribution of the seed photons, i.e.\ the source function, is
uniquely determined. The total luminosity $L_0$ of the seed photons thus obtained is also tabulated in
Table~\ref{table:result_column_model_B0_kT6keV} and \ref{table:result_column_model_B2_kT6keV}. The simulation results also allow the
luminosity $L_X$ to be determined.  Under the assumption that the accretion plasma itself is responsible for the X-ray emission and
reprocessing via Compton scattering, the luminosity $L_X$ obtained in the simulations has to agree with the observed luminosity
$L_\mathrm{obs}$, from which we constructed the simulation geometry, to keep self consistency of the model.

If this model were firmly established and the dependence on the
parameters were well understood, spectral fitting of the simulation
model to the observed spectra would be a powerful way to obtain the
physical properties of the accretion column.  Such analysis is,
however, beyond the scope of this paper.  Instead, we investigate the
basic behavior of the numerical model for different values of the
parameters. For example, the simulation results, presented in
\S\ref{subsec:magnetic_filed_effect}, show that for a fixed column
radius $r_0$ the Comptonized radiation becomes harder and more
luminous when $\xi$ increases.  Similarly, we have studied the
spectral dependence on $r_0$ with a fixed value of $\xi=1.5$. The
modeling data summarized in
Table~\ref{table:result_column_model_B2_kT6keV_radius} show that the
photon index $\Gamma$ becomes softer with increase of
$r_0$. Obviously, this softening is related to the decrease of the
optical depth across the column, which is approximately proportional
to $r_0{}^{-1}$.  The dependence of $L_X$ on $r_0$ is more complex.
On the one hand, the luminosity $L_0$ of the seed photons, which are produced via free-free
emissions, increases with $r_0$; however, the energy gain by the Comptonization
becomes effective for small $r_0$, i.e. for large optical
thickness.  Thus, the dependence of luminosity $L_X$ on $r_0$ is
determined by the competition of these two effects. The data, presented in
Table~\ref{table:result_column_model_B2_kT6keV_radius}, show
 that the luminosity barely depends on the column radius for the considered range of parameters.

\begin{table*}[htdp]
\caption{Fitted values of the simulated spectra for different column radii}
\begin{center}
\begin{tabular}{cccccccc}
\hline\hline
$r_0$ & $L_0$ & $A_1$ & $\Gamma$ & $A_2$ & $E_f$ & $L_X$ & $A_2/A_1$ \\
m & $\mathrm{erg\ s^{-1}}$ & $\mathrm{ph\ s^{-1}cm^{-2}keV^{-1}}$ & & $\mathrm{ph\ s^{-1}cm^{-2}keV^{-1}}$ & keV & $\mathrm{erg\ s^{-1}}$ & \\
\hline
200 & $2.26\times 10^{36}$ & $5.6\times 10^{-2}$ & $0.15$ & $3.0\times 10^{-4}$ & 6.7 & $3.8\times 10^{36}$ & $5.4\times 10^{-3}$ \\
300 & $3.05\times 10^{36}$ & $1.2\times 10^{-1}$ & $0.45$ & $3.4\times 10^{-4}$ & 6.6 & $4.0\times 10^{36}$ & $2.8\times 10^{-3}$ \\
400 & $3.67\times 10^{36}$& $1.9\times 10^{-1}$ & $0.58$ & $2.5\times 10^{-4}$ & 6.9 & $4.0\times 10^{36}$ & $1.3\times 10^{-3}$ \\
\hline
\end{tabular}
\end{center}
\label{table:result_column_model_B2_kT6keV_radius}
\end{table*}%

The inferred dependencies on the model parameters allowed us to find
acceptable self-consistent solutions for the observed luminosities,
which range from $1.5\times 10^{36}\ \mathrm{erg\ s^{-1}}$ to $6\times
10^{36}\ \mathrm{erg\ s^{-1}}$. The obtained parameters are tabulated
in Table~\ref{table:result_column_model_B2_kT6keV_self}, we note that
all acceptable solutions correspond to the unique value of $\xi$
parameter, $\xi=1.25$. 
This value is consistent with a rough estimate by Equation~\ref{eq:sonic_point} for typical photon energy of $\sim$5 keV, which should be close to the thermal energy. The cross sections for the different (perpendicular and parallel) directions can be written as $\sigma_\perp \simeq \sigma_\mathrm{T}$ and $\sigma_\parallel\simeq ({\bar E}/E_c)^2\sigma_\mathrm{T}$ (See Equations 6 and 7 of \citet{Becker:2007}).
As can be seen in
Table~\ref{table:result_column_model_B2_kT6keV_self}, the solutions
satisfy the self-consistent condition $L_\mathrm{obs}\sim L_X$ and
have reasonable values of the column radius.  These solution
reproduce the observed spectral hardening with increase of the luminosity, and
interestingly suggest a positive correlation between  the column radius and mass
accretion rate. However, at the present stage we cannot exclude presence of other
possible solutions.

\begin{table*}[htdp]
\caption{The self-consistent solutions of the accretion column spectrum}
\begin{center}
{\footnotesize
\begin{tabular}{ccccccccc}
\hline\hline
$L_\mathrm{obs}$ & $r_0$ & $L_0$ & $A_1$ & $\Gamma$ & $A_2$ & $E_f$ & $L_X$ & $A_2/A_1$ \\
$\mathrm{erg\ s^{-1}}$ & m & $\mathrm{erg\ s^{-1}}$ & $\mathrm{ph\ s^{-1}cm^{-2}keV^{-1}}$ & & $\mathrm{ph\ s^{-1}cm^{-2}keV^{-1}}$ & keV & $\mathrm{erg\ s^{-1}}$ & \\
\hline
$1.5\times 10^{36}$ &150 & $9.19\times 10^{35}$ & $3.4\times 10^{-2}$ & $0.58$ & $0.0$ & 17 & $1.2\times 10^{36}$ & $0$ \\
$3.0\times 10^{36}$ &150 & $1.68\times 10^{36}$ & $5.3\times 10^{-1}$ & $0.34$ & $2.9\times 10^{-4}$ & 6.6 & $3.0\times 10^{36}$ & $5.5\times 10^{-4}$ \\
$4.5\times 10^{36}$ &200 & $2.37\times 10^{36}$ & $5.5\times 10^{-1}$ & $0.17$ & $4.4\times 10^{-4}$ & 6.6 & $4.4\times 10^{36}$ & $8.0\times 10^{-4}$ \\
$6.0\times 10^{36}$ &300 & $3.51\times 10^{36}$ & $6.7\times 10^{-1}$ & $0.03$ & $4.4\times 10^{-4}$ & 6.9 & $6.2\times 10^{36}$ & $6.6\times 10^{-4}$ \\
\hline
\end{tabular}}
\end{center}
\label{table:result_column_model_B2_kT6keV_self}
\end{table*}%

Finally, we can interpret the spectral shape typical for a ``low state'' into which Vela X-1 sometimes drops.  Since the source still shows X-ray pulsations during the low states, it is evident that the accretion flow still reaches the magnetic poles of the neutron star.  In the low state, the accretion rate is an order of magnitude lower than that of the normal state, though the physical mechanism of the suppression is unknown.  The extremely soft power law with photon index of $\sim$2 is difficult to explain by thermal mechanism including the thermal Comptonization, as shown in Figure~\ref{fig:thermal_Comptonizaton_fit_parameters}.  Bulk Comptonization, however, naturally features such a soft spectrum.  We therefore suggest that the bulk Comptonization dominates over the thermal processes in the low state of the accreting neutron star.

\section{Conclusions}
\label{sec:conclusions}

We studied the process of Comptonization in the accretion column of binary pulsars. Our calculations were based on a Monte Carlo
framework MONACO and aimed to explain the data obtained with the \suzaku X-ray observatory from brightest wind-fed accreting neutron
star \vela.  The observational data used were reported in \citetalias{Odaka:2013}, where a long exposure (100~ks) was split into 56 short timescale (2~ks) spectra to study the variability.
These X-ray spectra are best fitted  by the NPEX model, which is a combination of two different-slope
power laws with a common exponential cutoff, with an obvious cyclotron resonance scattering feature at 50 keV \citepalias{Odaka:2013}. 

Our study has shown that NPEX-type spectra are consistent with a Comptonization origin. This association allows us to link
phenomenological parameters in the NPEX model with the physical properties of the production site of the X-ray emission. In particular, we show that the
photon index $\Gamma$ can be used as a good measure of the source optical depth. This therefore suggests a physical interpretation for an observed
correlation between the photon index $\Gamma$ and X-ray luminosity: the optical thickness of the accreted plasma increases with the mass
accretion rate onto the neutron star.

A more detailed study of a magnetized accretion column showed that the broad-band X-ray spectra obtained with \suzaku can be well explained as Comptonized
emission produced in the accretion column with a strong magnetic field of $\sim 10^{12}$ G. Importantly, within this interpretation, the \suzaku spectra are consistent with  physically reasonable values of the parameters describing the geometry of the accretion column.

Finally, we find that thermal Comptonization is unlikely to be responsible for the soft spectra (with a photon index of $\sim$2) measured during  ``low states'', in which the source luminosity becomes an order of magnitude lower than that of the normal state. Since these states still shows X-ray pulsations, indicating that the accretion flow reaches to the magnetic poles on the neutron star \citepalias{Odaka:2013}, we suggest that such a soft power law is most likely produced via Comptonization dominated by the bulk motion rather than thermal processes.

\acknowledgments

The authors are grateful to Prof.\ Chris Done for her useful comments on the manuscript. H.~Odaka and Y.~Tanaka had been supported by research fellowships of the Japan Society for the Promotion of Science for Young Scientists. This work was supported by JSPS KAKENHI Grant Number 24740190. This work was supported in part by Global COE Program (Global Center of Excellence for Physical Sciences Frontier), MEXT, Japan.

\clearpage

\appendix

\section{Calculation of Thermal and Bulk Comptonization}\label{sec:calculation}

Comptonization is one of the main processes for production of X-ray radiations in accretion flows.
In this process a photon is up-scattered to higher energy by a high-energy electron. Depending on energy distribution of the electrons, the dominant contribution to the Comptonization can be related either to thermal or bulk motion of the electrons. Therefore, one distinguishes {\it thermal} and {\it bulk} Comptonization processes. In compact binary systems, both of these processes typically lead to formation of spectra peaking in the X-ray energy band. Indeed, in the case of thermal Comptonization, the energy of up-scattered photons saturates close to the electron temperature. Similarly, free fall velocity, which serves as a good measure of the bulk velocity, onto a neutron star achieves $0.5c$. Thus, bulk Comptonization is also expected to peak in the X-ray band.

Many important features of the Comptonization have been obtained with an analytical approach. In particular, Kompaneets Equation proved to be very useful for studying thermal Comptonization  \citep[e.g.,][]{Rybicki:radiative_process,Sunyaev:1980, Titarchuk:1994}. However, the analytical approaches have several limitations \citep[see, e.g.,][]{Hua:1997}. For example, the photon transport equation typically implies that the energy and position of photons are continuous functions of time. This assumption is a good approximation for low energy bands since scattering at low energy does not change the photon energy significantly.
At higher energies, this assumption breaks due to a substantial change of the photon energy in a single scattering.  The analytical methods are also not suitable for a precise description of the photon escape, which is strongly affected by the geometrical structure of the production region. By contrast, it is possible to address all these limitations in Monte-Carlo-type simulations, where both detailed photon transport and precise scattering process  are automatically accounted for. While Monte Carlo simulations  have been performed  by a few groups \citep[see, e.g.,][]{Pozdnyakov:1983, Hua:1997} for thermal Comptonization, we exploit an approach which includes both thermal and bulk Comptonization in complicated geometries.

\begin{figure}[tbp]
\begin{center}
\includegraphics[width=8.5cm]{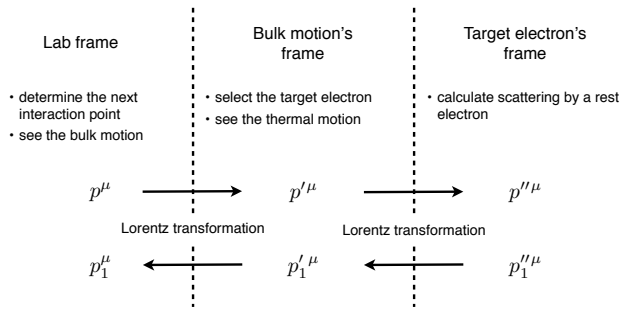}
\caption{Schematic outline of calculation for inverse Compton scattering. $p^\mu$ and $p_1^\mu$ denote energy-momentum vectors of an incident and scattered photon, respectively. ($\mu$ is a coordinate index that runs from 0 to 3.)}
\label{fig:calc_compton}
\end{center}
\end{figure}

We build our numerical procedure on a sequence of computations performed in different coordinate systems with the physical quantities being related via corresponding Lorentz transformations. We use three coordinate systems: {\it laboratory} (i.e., the observer system), {\it bulk} (where electron's bulk velocity is zero) and {\it electron rest} frames. A sketch of the implementation is shown in figure~\ref{fig:calc_compton}. The laboratory frame is used to compute the photon path, and the electron rest frame  is suitable for detailed calculations of the scattering. 

The mean interaction rate for a photon in the laboratory frame can be obtained by averaging the number of scattering ($\nu$) per unit of time ($ds=cdt$) over the electron distribution:
\begin{equation}\label{eq:averaged_interaction_rate}
\left\langle \dfrac{d\nu}{ds} \right\rangle 
= \int dE_\mathrm{e} \int d\bm{\Omega}_\mathrm{e} (1-\beta\cos\theta)\sigma(E')N_\mathrm{e}(E_\mathrm{e}, \bm{\Omega}_\mathrm{e}),
\end{equation}
where $\beta=v_{\rm e}/c$ is an electron velocity; $\theta$ is the angle between the electron and photon velocities;  $N_\mathrm{e}(E_\mathrm{e},\bm{\Omega}_\mathrm{e})=dN/(dVdE_\mathrm{e}d\bm{\Omega}_\mathrm{e})$ is the electron distribution function for energy $E_\mathrm{e}$ and direction $\bm{\Omega}_\mathrm{e}$; $\sigma (E')$ is the invariant scattering cross section computed in the electron rest frame \citep[for detail see, e.g.,][]{Landau:classical_field}. Since the physical region of interest here has an X-ray energy $E\lesssim100$ keV and a low electron temperature $kT\lesssim 10$ keV, the cross section has weak energy dependence, and the thermal motion of the electrons is negligible in the calculation of the mean free path. Thus, 
Equation \ref{eq:averaged_interaction_rate} can be simplified as
\begin{equation}
\left\langle \dfrac{d\nu}{ds} \right\rangle \approx (1-\beta_\mathrm{b}\cos\theta)\sigma(E')n_\mathrm{e}.
\end{equation}
Here,  $\beta_{\rm b}=v_{\rm b}/c$ is the bulk velocity; $\theta$ is the angle between the bulk motion and the incident photon direction; $E'$ is the Doppler shifted energy of the incident photon in the bulk motion reference frame. Consequently, the mean free path of photons is given by
\begin{equation}
l  \approx \frac{1}{(1-\beta_\mathrm{b}\cos\theta)\sigma(E')n_\mathrm{e}}.
\end{equation}

Once the Monte Carlo code samples the interaction point based on the photon direction and the value of the mean free path $l$, the scattering process is computed. The numerical procedure for calculation of the photon-electron interaction consists of two steps: first, the scattering electron is selected via evaluation in the bulk reference frame; second, the final state of the scattered photon is computed in the electron rest frame.
Using Equation (\ref{eq:averaged_interaction_rate}) in the bulk reference frame, we can evaluate the interaction probability as
\begin{equation}
P(E'_\mathrm{e}, \bm{\Omega}'_\mathrm{e}) \propto (1-\beta'\cos\theta')\sigma(E'')N_\mathrm{e}(E'_\mathrm{e}).
\end{equation}
Here, prime indicates values in the reference frame of the bulk motion, not the laboratory frame; double prime indicates values in the electron rest frame. Since the energy dependence of the cross section is negligible for low temperature ($kT\lesssim10$ keV) plasma, the direction and energy of the electron can be sampled independently.
Namely, the electron velocity $v'=\beta'/c=\sqrt{2E'_\mathrm{e}/m_\mathrm{e}}$ is sampled from the Maxwell distribution
\begin{equation}
N_\mathrm{e}(v)dv=\sqrt{\dfrac{2}{\pi}\left(\dfrac{m_\mathrm{e}}{kT}\right)^3}v^2 \exp\left(-\dfrac{m_\mathrm{e}v^2}{2kT}\right)dv\,;
\end{equation}
and the direction of the electron motion  $\beta'$ obtained from
\begin{equation}
P(\cos\theta') \propto 1-\beta'\cos\theta'.
\end{equation}
The numerical procedure outlined above allows us to obtain energy $E''_1$ and direction $\bm{\Omega}''_1$ of the scattered photon in the electron rest frame by using the Klein-Nishina differential cross section.
Finally, we use a reverse Lorentz transformation in order to obtain the result of the Compton scattering seen by the observer.


In order to verify our simulation code, we calculated thermal Comptonization radiation (the bulk velocity was assumed to be zero) from a
spherically symmetric cloud of fully ionized plasma.  This is the simplest model to check X-ray spectra generated by Comptonization.
We have performed calculations for two different electron temperatures of $kT=0.64$ keV and 6.4 keV.

\begin{figure*}[tbp]
\begin{center}
\includegraphics[width=7cm]{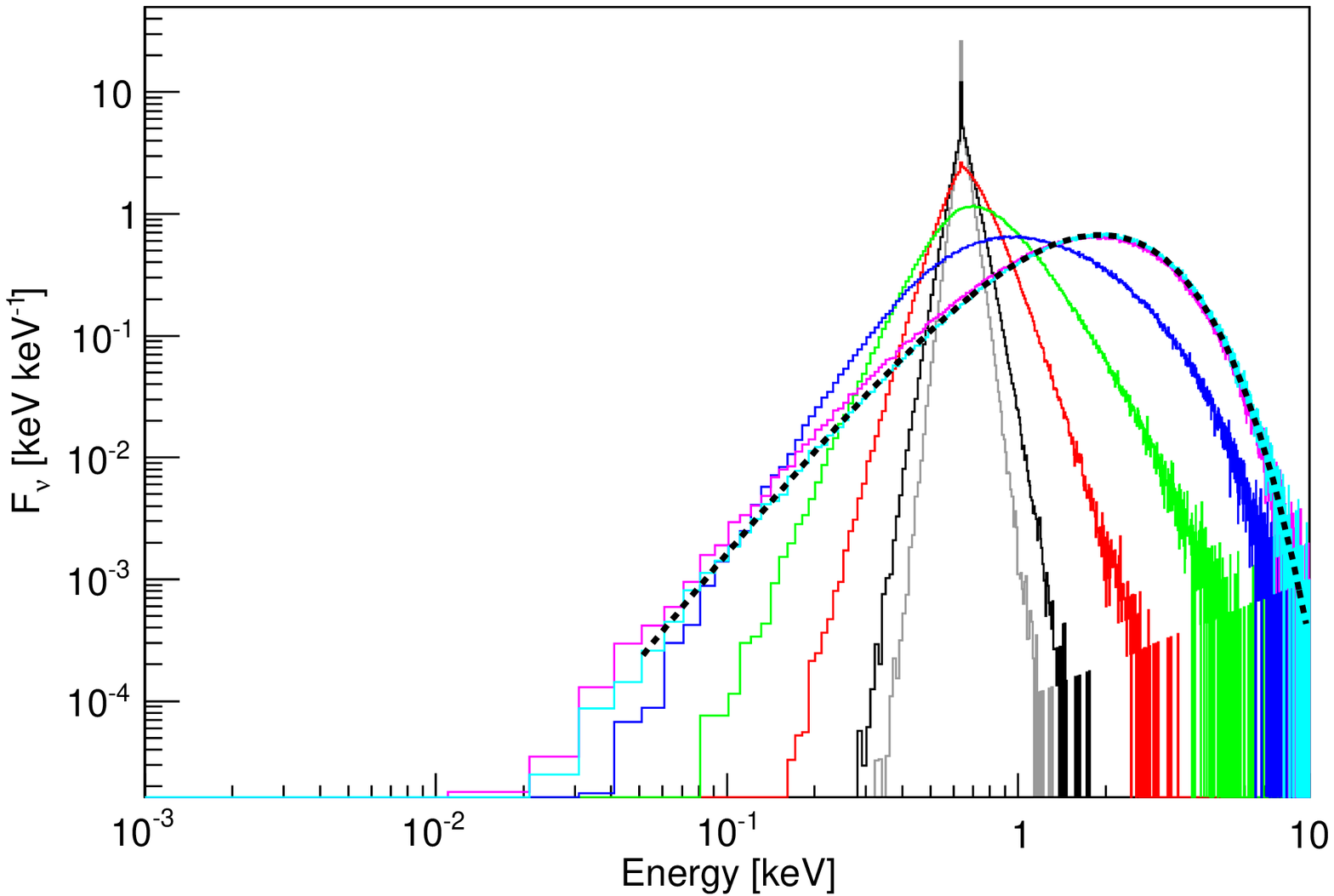}
\includegraphics[width=7cm]{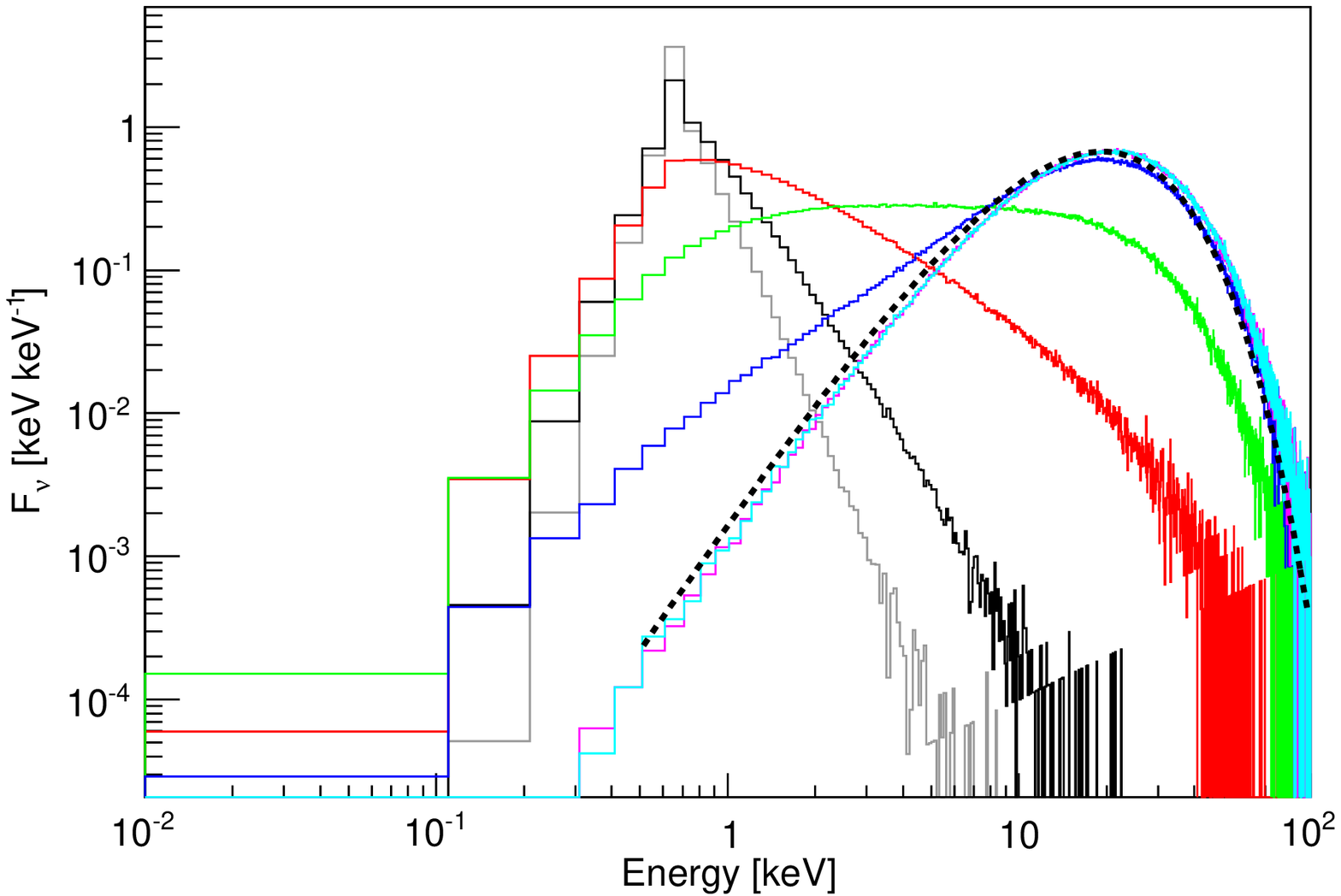}
\caption{Right: energy spectra $F_\nu$ of Comptonized radiations from a spherical plasma with a temperature of $kT=0.64$ keV for different radial optical (Thomson) depths $\tau=1$ (grey), 2 (black), 5 (red), 10 (green), 20 (blue), 50 (magenta), and 100 (cyan). The source photon is generated at the center and $E_0=0.64$ keV.  The Wien spectrum at the same temperature is superposed as the dashed curve. Left: the same as the right panel but for an electron temperature of $kT=6.4$ keV.}
\label{fig:spec_thermal_compton}
\end{center}
\end{figure*}

The initial photons start at the center of the cloud $\bm{x}_0=(0,0,0)$ with an initial energy 0.64 keV.
The left panel of Figure \ref{fig:spec_thermal_compton} shows the results for different radial optical (Thomson) depths ranging from $\tau=1$ to 100 and for the common temperature $kT=0.64$ keV.
For $\tau=1$, the spectrum is not largely altered from the initial spectrum because the escaped photons have experienced zero or a few scatterings by electrons.
As the optical depth increases, the photon spectrum ``evolves'' to the saturated Comptonization or thermal equilibrium.
The spectrum for $\tau=100$, which is a result of $\sim$10000 interactions with electrons, nicely agrees with the Wien spectrum at $kT=0.64$ keV, as predicted theoretically.

The right panel of Figure \ref{fig:spec_thermal_compton} shows the same calculation except for the electron temperature $kT=6.4$ keV, i.e., ten times higher than the first example.
The initial conditions or source photons are the same for both the models.
The photons undergo faster heating due to the higher electron temperature.
There is a broad peak at $E\simeq 3kT\sim 20$ keV, the ``Wien hump'' generated by the saturated Comptonization for a large optical depth, whereas the unsaturated emission from $\tau=1$ forms a power-law-like spectrum.
The spectral index of the power-law component strongly depends on the optical depth of the Comptonizing cloud.

\bibliographystyle{apj}
\bibliography{velax1_suzaku}




\end{document}